\documentclass{article}

\usepackage{arxiv}

\usepackage[utf8]{inputenc} 
\usepackage{url}            
\usepackage{booktabs}       
\usepackage{amsfonts}       
\usepackage{nicefrac}       
\usepackage{microtype}      

\usepackage[T1]{fontenc}
\usepackage{lmodern}
\usepackage{graphicx}
\usepackage{multirow}
\usepackage{bbm}
\usepackage{amssymb,amsmath}
\usepackage{mathtools}
\usepackage{booktabs}	
\usepackage{float}	
\usepackage{subcaption} 
\usepackage{enumitem}
\usepackage{bm}
\usepackage{color,soul} 
\usepackage{array}
\usepackage{siunitx} 
\usepackage[normalem]{ulem} 
\usepackage{comment} 
\usepackage{tikz} 
\usetikzlibrary{shapes, arrows, calc}
\usepackage[colorlinks,citecolor=green,urlcolor=blue,
bookmarks=false,hypertexnames=true]{hyperref}

\title{On spike-and-slab priors for Bayesian equation discovery of nonlinear dynamical systems via sparse linear regression}

\author{
  Rajdip Nayek \\
  Department of Mechanical Engineering\\
  University of Sheffield\\
  Mappin Street, Sheffield S1 3JD\\
  \texttt{r.nayek@sheffield.ac.uk} \\
   \And
 Ramon Fuentes \\
 Callsign Inc.\\
 London \\
 \And
 Keith Worden\\
 Department of Mechanical Engineering\\
 University of Sheffield\\
 Mappin Street, Sheffield S1 3JD\\
 \texttt{k.worden@sheffield.ac.uk} \\
 \And
 Elizabeth J.~Cross\\
 Department of Mechanical Engineering\\
 University of Sheffield\\
 Mappin Street, Sheffield S1 3JD\\
 \texttt{e.j.cross@sheffield.ac.uk} 
}

\date{}

\newcommand{\norm}[1]{\left\lVert#1\right\rVert}
\newcommand{\ra}[1]{\renewcommand{\arraystretch}{#1}}
\newcommand{\abs}[1]{\left\lvert#1\right\rvert}
\newcommand{\matr}[1]{\mathbf{{#1}}}    
\newcommand{\ve}[1]{\bm{{#1}}}          
\newcommand{\x}{\ve{x}}
\newcommand{\z}{\ve{z}}
\newcommand{\y}{\ve{y}}

\newcommand{\pz}{p_0}
\newcommand{\vs}{v_s}

\newcommand{\f}{\mathbf{f}}
\newcommand{\uinp}{\ve{u}}
\newcommand{\err}{\ve{\epsilon}}

\newcommand{\mD}{\matr{D}}
\newcommand{\vtheta}{\ve{\theta}}
\newcommand{\zeros}{\matr{0}}
\newcommand{\eye}{\matr{I}}
\newcommand{\ones}{\ve{1}}
\newcommand{\brc}[1]{\left(#1\right)}       
\newcommand{\cbk}[1]{\left\{#1\right\}}     
\newcommand{\sbk}[1]{\left[#1\right]}       

\newcommand{\Dr}[1]{\matr{D}_{r}^{#1}}
\newcommand{\muz}{\ve{\mu}}
\newcommand{\Sigmaz}{\matr{\Sigma}}

\newcommand{\ND}[1]{\mathcal{N}\brc{#1}}
\newcommand{\IG}[1]{\mathcal{IG}\brc{#1}}
\newcommand{\pr}[1]{p\brc{#1}}
\newcommand{\F}{\mathcal{M}}
\newcommand{\R}{\mathbb{R}}

\newcommand{\A}[1]{\matr{A}_{#1}}

\begin{document}
\maketitle

\begin{abstract}
This paper presents the use of spike-and-slab (SS) priors for discovering governing differential equations of motion of nonlinear structural dynamic systems. The problem of discovering governing equations is cast as that of selecting \textit{relevant} variables from a predetermined dictionary of basis variables and solved via sparse Bayesian linear regression. The SS priors, which belong to a class of discrete-mixture priors and are known for their strong sparsifying (or \textit{shrinkage}) properties, are employed to induce sparse solutions and select relevant variables. Three different variants of SS priors are explored for performing Bayesian equation discovery.	
As the posteriors with SS priors are analytically intractable, a Markov chain Monte Carlo (MCMC)-based Gibbs sampler is employed for drawing posterior samples of the model parameters; the posterior samples are used for variable selection and parameter estimation in equation discovery. The proposed algorithm has been applied to four systems of engineering interest, which include a baseline linear system, and systems with cubic stiffness, quadratic viscous damping, and Coulomb damping. The results demonstrate the effectiveness of the SS priors in identifying the presence and type of nonlinearity in the system. Additionally, comparisons with the Relevance Vector Machine (RVM) -- that uses a Student's-$t$ prior -- indicate that the SS priors can achieve better model selection consistency, reduce false discoveries, and derive models that have superior predictive accuracy. Finally, the Silverbox experimental benchmark is used to validate the proposed methodology.
\end{abstract}


\section{Introduction} 
Spurred on by the rapid increase in computational power and growing rates of data collection, recent years have seen a booming interest in discovering governing differential equations of motion of nonlinear dynamical systems from time-series data \cite{bongard2007automated, schmidt2009distilling, brunton2016discovering, mangan2016inferring, schaeffer2017sparse, mangan2017model, kaiser2018sparse, champion2019data, schaeffer2020extracting, boninsegna2018sparse, mangan2019model, stender2019recovery, rudy2017data, zhang2018robust, chen2020deep, rudy2019deep,raissi2018multistep,raissi2018deep}.
Governing differential equations of motion are ordinary or partial differential equations that characterise system behaviour and provide an understanding of the physics of the underlying phenomena.  
Historically, such equations have been derived based on first principles and some prior knowledge of the nature of the system; once known, they can be used for further analysis, prediction and control of the system. In structural dynamics, the governing equations of motion can often be represented as a state-space model (SSM) of the form \footnote{Note the explicit time dependence of the states $\x(t)$ and inputs $\uinp(t)$ has been suppressed to simplify notation.}, 
\begin{align} \label{eq:ssform}
	\dot{\x} = \F (\x) + \uinp 
\end{align}
where $\x$ is the state vector of system responses, $\dot{\x}$ is the time derivative of the state vector, $\F$ is the function of states $\x$ embedding the equation of motion of the structure, and $\uinp$ is the vector of external input forces that are assumed to enter linearly in Eq.~\eqref{eq:ssform}. 
This form in equation is quite common in structural dynamics, although a more general form of Eq.~\eqref{eq:ssform} would replace $\F$ with a function of both the state and the input vectors. 
Due to the ubiquitous presence of nonlinearity in modern structural systems, the structural equations of motion, represented by the model $\F$, typically includes a number of nonlinear terms in $\x$. 
However, in most situations, the true form of $\F$ is unknown, and hence, there arises a need to recover the model $\F$, i.e.\ the underlying equations of motion. 
Formally, the task of recovering $\F$ involves solving two sub-problems: \textit{model selection}, which aims to identify a suitable form of $\F$, and \textit{parameter estimation}, which determines the unknown parameters of the chosen form of $\F$. Individually, both these problems have received significant attention in the remit of structural dynamics as well as in the broader context of nonlinear system identification, and the interested reader can find excellent review papers \cite{kerschen2006past, noel2017nonlinear} on nonlinear system identification. 


When the goal is to discover a parametric form of $\F$, the identified model needs to satisfy two essential attributes: (a) good \textit{prediction power}, by the virtue of which it is able to predict future observations effectively without
suffering from over-fitting, and (b) \textit{interpretability}, so that the model includes only a few features (or predictors) that exhibit the strongest effect, thus providing a better understanding of the underlying process. Typically, when prediction is the only aim, the actual choice of the features is of less interest, as long as the fit to the data is good. However, when the aim is also to understand the physical phenomena generating the responses -- which is the case here -- there is a need to search for the real but unknown relationship between the responses and the features. For a good understanding of the relationship, it is important to select only the relevant features i.e., the features that matter.

Traditional model selection procedures work by postulating a small set of \textit{interpretable} models -- chosen based on expert intuition and domain knowledge. The ``best'' model is selected as the one that achieves a desired balance of model complexity and goodness-of-fit, judged by some information-theoretic criteria such as AIC \cite{akaike1974new}, BIC \cite{schwarz1978estimating}, etc. Nonetheless, these traditional procedures can become prohibitive when prior knowledge is limited and the number of candidate models is large (in the order of hundreds or greater). With the rapid advancement in data-driven modelling in the last two decades, there has been an emergence of alternative frameworks of model selection that rely less on expert knowledge and more on data. An early effort towards data-driven modelling for equation discovery was symbolic regression \cite{bongard2007automated, schmidt2009distilling}, which searches through a library (or \textit{dictionary}) of simple and interpretable basis variables to identify the parametric form of the governing equations of a nonlinear dynamical system. While this approach works well for discovering interpretable physical models, its dependence on evolutionary optimisation for selecting the \textit{relevant} variables from the dictionary makes it computationally expensive, and unsuited to large-scale problems. In a more recent study \cite{brunton2016discovering}, the model discovery process was reformulated in terms of sparse linear regression, which makes the variable selection process amenable to solution using efficient sparsity-promoting algorithms, thus providing a computationally-cheaper alternative. Since then, the sparse regression approach for data-driven equation discovery of differential equations has been further developed in many studies. Examples include sparse identification of biological networks with rational basis variables \cite{mangan2016inferring}, model selection using an integral formulation of the differential equation to reduce noise effects \cite{schaeffer2017sparse}, model selection for dynamical selection combining sparse regression and information criteria \cite{mangan2017model}, extension of sparse identification to nonlinear systems with control \cite{kaiser2018sparse}, discovery of coordinates for sparse representation of governing equations \cite{champion2019data}, extracting structured differential equations with under-sampled data \cite{schaeffer2020extracting}, sparse learning of stochastic dynamical equations \cite{boninsegna2018sparse}, model selection for nonlinear dynamical systems with switching behaviour \cite{mangan2019model}, recovery of differential equations from short impulse response time-series data \cite{stender2019recovery}, identification of parametric partial differential equations \cite{rudy2017data, zhang2018robust, chen2020deep}. There are also studies that proposed \textit{black-box} approaches using deep neural networks \cite{rudy2019deep,raissi2018multistep,raissi2018deep} for equation discovery of differential equations; however, they are mostly useful for forecasting and do not provide explicit equations for interpretation.

The work presented in this paper adopts the sparse-regression-based parametric-equation-discovery approach for recovering the governing equation of motion of a structural dynamic system. In this approach, it is assumed that the function $\F$ consists of only a few terms, making it sparse in the space of possible functions. The assumption is generally true for many systems of engineering interest, as their governing physics is often simple and interpretable. To recover $\F$, the idea is to first express $\F$ as a weighted linear combination of a large number of simple intepretable basis functions (or basis variables), and then apply algorithms to select a \textit{relevant} subset of variables that best explains the measurements. 
As such, in this approach, the model selection problem is turned into a variable selection problem.

\begin{figure}[h]
	\begin{center}
		\includegraphics[scale = 0.5]{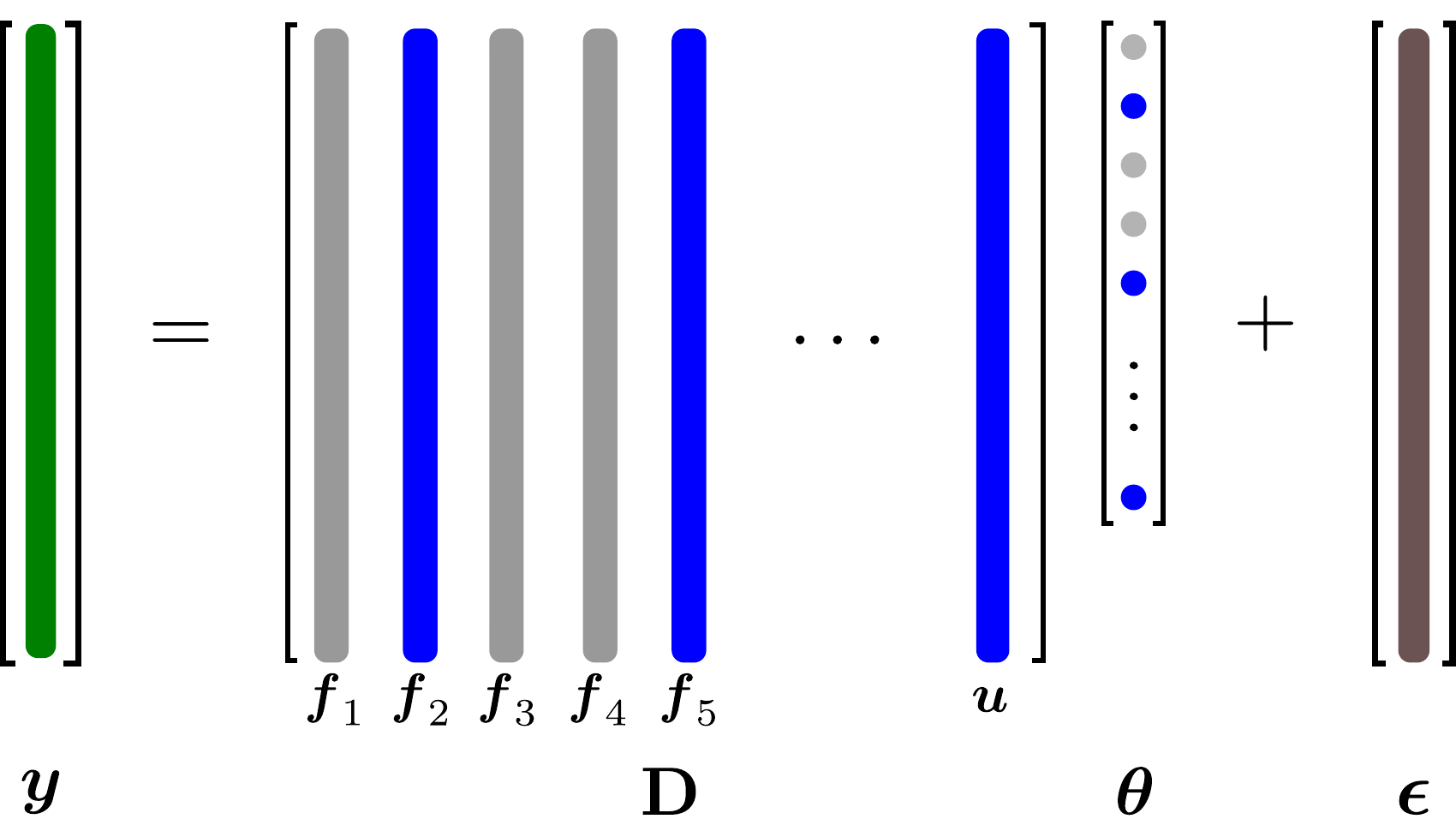}
		\caption{Sparse linear regression for selection of relevant basis variables (shown in blue) in parametric equation discovery approach.}
		\label{f:SLR}
	\end{center}
\end{figure}

To elaborate on the procedure, consider the example of a Single Degree-of-Freedom (SDOF) oscillator with equation of motion of the form,
\begin{align} \label{eq:sdof}
	m \ddot{q} + c \dot{q} + k q + g \brc{q,\dot{q}} = u
\end{align}
where $m$, $c$, $k$ are the mass, damping, and stiffness, $g$ is an arbitrary nonlinear function of displacement $q$ and velocity $\dot{q}$; $\ddot{q}$ is the acceleration, and $u$ is the input forcing function. An SSM for this system can be written as,
\begin{align}
	\dot{x}_1 &= x_2 \label{eq:first_eqn}\\
	\dot{x}_2 &= \frac{1}{m} \brc{u - k x_1 - c x_2 - g(x_1, x_2)} \label{eq:main_eqn}
\end{align}
with $x_1 = q$ and $x_2 = \dot{q}$. 
Eq.~\eqref{eq:first_eqn} can be ignored as it simply provides the definition of velocity; Eq.~\eqref{eq:main_eqn} captures the governing equation of the structure's motion. To uncover the underlying structure of the right hand side of Eq.~\eqref{eq:main_eqn}, a large dictionary of basis variables $f_1(x_1,x_2), f_2(x_1,x_2),\ldots, f_l(x_1,x_2)$ is constructed, containing several functional forms such as polynomial terms, trigonometric terms, etc. The left-hand side, which represents acceleration $\dot{x}_2$, is then expressed on the right as a weighted linear combination of the basis variables of the dictionary, 
\begin{align}
	\dot{x}_2 &\approx \theta_1 f_1(x_1, x_2) + \theta_2 f_2(x_1, x_2) + \cdots + \theta_l f_l(x_1, x_2) + \theta_{l+1} u
\end{align}
where $\cbk{\theta_1, \theta_2, \cdots, \theta_l, \theta_{l+1}}$ are the associated weights. Note that the input is also added to the dictionary to identify its corresponding weight. Given noisy time-series measurements $\cbk{x_{1,j}, x_{2,j}, \dot{x}_{2,j}, u_j}_{j=1}^N$, where $j$ in the subscript indicates time point $t_j$, the above problem reduces to a linear regression problem,
\begin{align} \label{eq:linreg_expand}
	\underbrace{\begin{bmatrix}
			\dot{x}_{2,1} \\
			\dot{x}_{2,2} \\
			\vdots \\
			\dot{x}_{2,N}
	\end{bmatrix}}_{\y}
	&= 
	\underbrace{\begin{bmatrix}
			f_1 \brc{x_{1,1}, x_{2,1}} &  f_2 \brc{x_{1,1}, x_{2,1}}  & \cdots & f_l \brc{x_{1,1}, x_{2,1}} & u_1   \\
			f_1 \brc{x_{1,2}, x_{2,2}} &  f_2 \brc{x_{1,2}, x_{2,2}}  & \cdots & f_l \brc{x_{1,2}, x_{2,2}} & u_2   \\ 
			\vdots                   &  \vdots                    & \ddots & \vdots                   & \vdots \\
			f_1 \brc{x_{1,N}, x_{2,N}} &  f_2 \brc{x_{1,N}, x_{2,N}}  & \cdots & f_l \brc{x_{1,N}, x_{2,N}} & u_N    
	\end{bmatrix}}_{\mD}
	\underbrace{\begin{bmatrix}
			\theta_1 \\ \theta_2 \\ \vdots \\ \theta_l \\ \theta_{l+1}
	\end{bmatrix}}_{\vtheta}
	+ 
	\underbrace{\begin{bmatrix}
			\epsilon_1 \\ \epsilon_2 \\ \vdots \\ \epsilon_N 
	\end{bmatrix}}_{\err} 
\end{align}
which can be written in a compact matrix-vector notation as,
\begin{align} \label{eq:linreg}
	\y &= \mD \vtheta + \err
\end{align}
Here, $\y \in \R^{N \times 1}$ is a vector of observations of acceleration, $\mD \in \R^{N \times P}$ is a dictionary\footnote{The number of columns in the dictionary has been redefined as $P = l+1$} matrix composed using states (i.e.\ displacement and velocity) and input force, $\vtheta \in \R^{P \times 1}$ is the vector of basis weights and $\err  \in \R^{N \times 1}$ is the residual error vector taking into account model inadequacies and measurement errors. The task is now to select which basis variables from the dictionary are to be included in the final estimated model $\hat{\F}$. As only a few basis variables from the dictionary are assumed to contribute actively to the governing dynamics, the solution of $\vtheta$ would be sparse, i.e.\ would have only a few weights that are significantly different than zero; hence, it is reasonable to seek sparse solutions of $\vtheta$ in the above linear regression problem, as illustrated in Figure~\ref{f:SLR}.

Classical penalisation methods \cite{hastie2015statistical}, including lasso, ridge, and elastic net penalties, can offer sparse solutions to the linear regression problem. They are deterministic approaches that employ constrained optimisation schemes to achieve sparse solutions, in that they add a convex penalty function to the usual least-squares objective and shrink the small weights to zero while leaving out a few large weights. Another popular deterministic method seeking sparse solutions to the linear regression problem is the sequential threshold least-squares algorithm, which iteratively solves the least-squares problem while zeroing out the small weights in successive iterations. This algorithm has been used in many works on equation discovery of nonlinear dynamical systems, including \cite{brunton2016discovering}. However, a common drawback of the deterministic approaches is that the results are sensitive to the choice of a regularisation parameter, and its tuning is required externally by cross-validation. 

In this work, a sparse Bayesian learning approach \cite{tipping2001sparse, wipf2004sparse} is adopted over a deterministic penalisation or thresholding approach, to solve the sparse linear regression problem. Apart from the usual advantage of uncertainty quantification, the sparse Bayesian learning framework offers three additional advantages: (a) it allows for natural penalisation through prior distributions, (b) the penalty parameter is simultaneously estimated with other model parameters and does not require determination through cross-validation, and (c) Bayesian techniques using Markov Chain Monte Carlo (MCMC) sampling facilitate a more straightforward implementation of non-convex penalty functions, unlike classical approaches which use convex penalty functions to achieve a unique minimum.  


In sparse Bayesian learning, sparsity is induced by placing sparsity-promoting (or \textit{shrinkage}) priors on the weights. These priors tend to shrink small weights to zero while allowing a few large weights to escape shrinkage. The densities of these priors feature a strong peak at zero and heavy tails: the peak at zero enforces most of the values to be (near) zero while heavy tails allow a few non-zero values. This structure of the priors tends to produce a selective shrinkage of the weights of the linear model, i.e.\ the posterior distributions of most weights are shrunk towards zero while a small set of weights have a large probability of being significantly different from zero \cite{seeger2010optimization}. Examples of such priors include: Laplace \cite{seeger2008bayesian}, Student's-$t$ \cite{tipping2001sparse}, Horseshoe \cite{carvalho2009handling}, and spike-and-slab \cite{mitchell1988bayesian, george1993variable, geweke1996variable, george1997approaches, ishwaran2005spike}. An overview of various shrinkage priors used in sparse Bayesian linear regression can be found in \cite{polson2010shrink, van2019shrinkage}.

The use of sparse Bayesian learning in data-driven equation discovery has been explored only very recently, and the current state of research in this direction has been quite limited. A handful of research that exists has mostly focussed on obtaining sparse solutions via a particular implementation of the Student's-$t$ prior -- the Relevance Vector Machine (RVM) \cite{tipping2001sparse}. Unlike common Bayesian algorithms that use MCMC-based random sampling, the RVM performs a marginal likelihood optimisation to yield parameter posteriors. The RVM was used in \cite{fuentes2019efficient} for equation discovery of nonlinear structural dynamic systems. A magnitude-based weight-thresholding was combined with RVM in \cite{zhang2018robust} for discovery of governing partial differential equations. Recently, \cite{zhang2019robust} extended the hybrid algorithm with a subsampling approach to discover governing equations of nonlinear systems in the presence of outliers.

\begin{figure}[h]
	\begin{center}
		\begin{subfigure}{.42\textwidth}
			\centering
			\includegraphics[scale = 0.5]{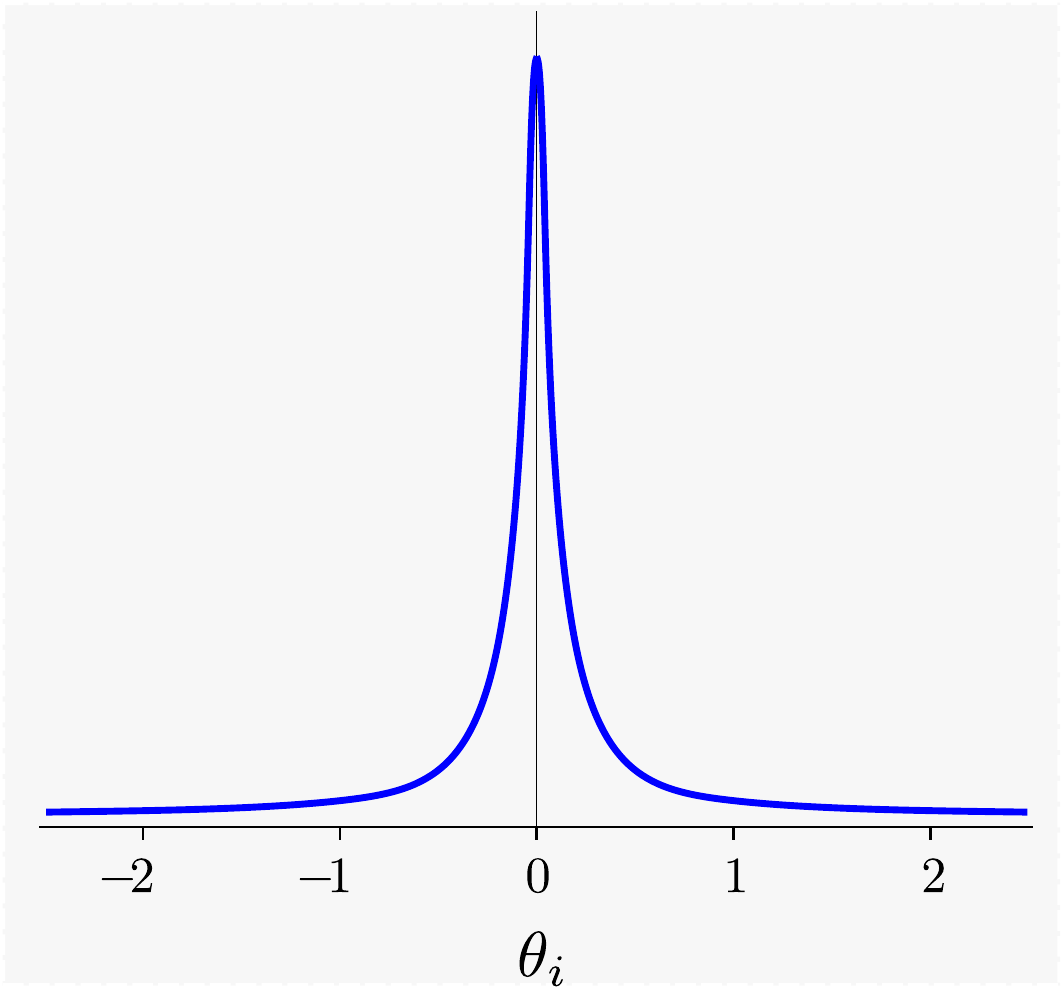}  
			\caption{Student's-$t$}
			\label{fig:sub-first}
		\end{subfigure}
		\begin{subfigure}{.42\textwidth}
			\centering
			\includegraphics[scale = 0.5]{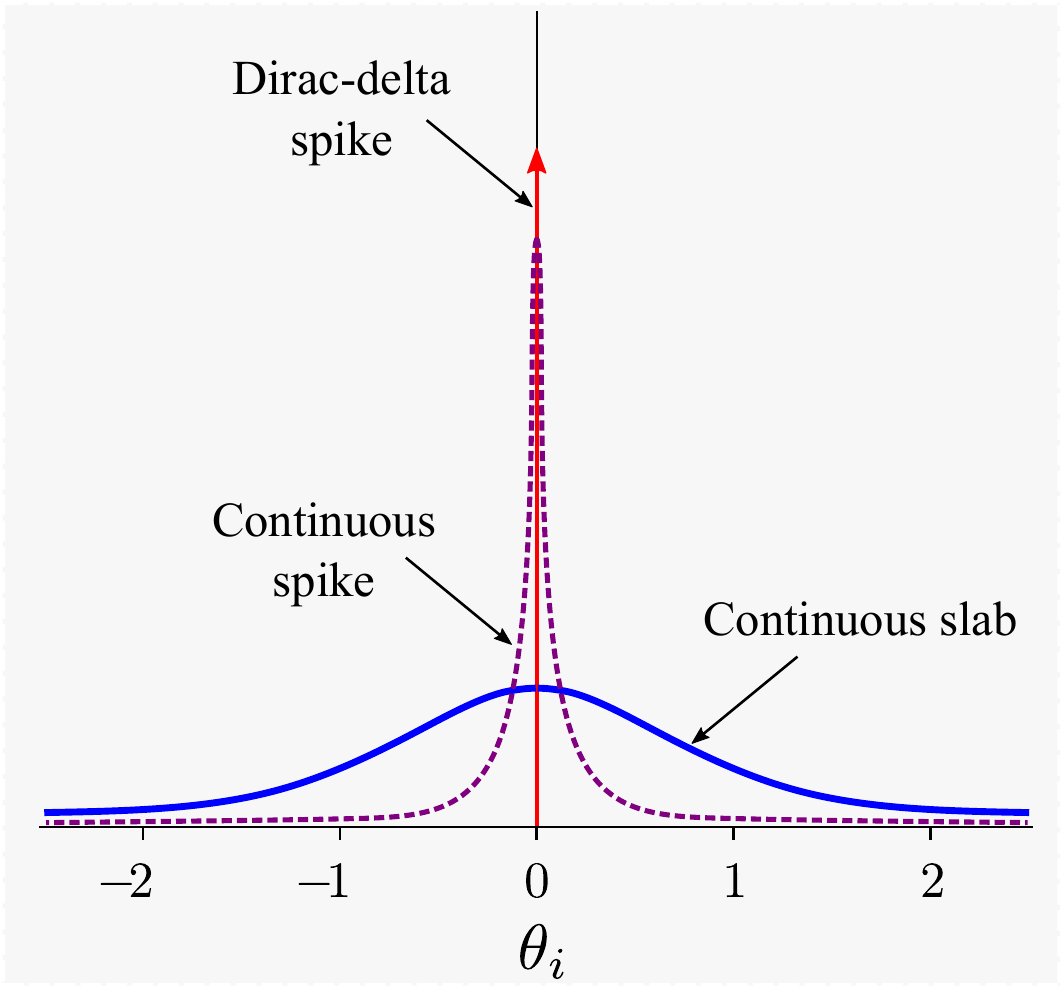}  
			\caption{Spike-and-slab}
			\label{fig:sub-second}
		\end{subfigure}
		\caption{Probability density functions of (a) the Student's-$t$ prior, and (b) the spike-and-slab (SS) prior with either a Dirac-delta spike (displayed by an arrow pointing upwards) or a narrow continuous spike (displayed by dotted line).}
		\label{f:StudT_vs_SS}
	\end{center}
\end{figure}

The critical challenge in the equation discovery approach is to learn the correct set of basis variables from the dictionary $\mD$. Although the RVM can provide quick results, it is based on the Student's-$t$ prior that has less selective shrinkage capabilities compared to priors such as the SS prior. Figure~\ref{f:StudT_vs_SS} provides a visual illustration of the densities of the Student's-$t$ and SS priors. The Student's-$t$ prior is not as peaked around zero, hence it allows some weights -- which should truly be zero -- to take non-zero values. In an equation discovery setting, this issue may lead to more terms being selected than is true and may hinder the interpretability of the learned model. On the other hand, an SS prior comprises a small (or a point) mass at zero (the \textit{spike}) for small weights, and a diffused density (the \textit{slab}) for the large weights. The spike is capable of shrinking the small coefficients towards zero; hence the SS prior can induce stronger selective shrinkage of the coefficients compared to the Student's-$t$ prior. Previously, the authors proposed the use of SS priors in equation discovery of nonlinear systems \cite{nayek2020spike}, on which the current work builds. 

This paper explores the performance of three different variants of SS priors in Bayesian equation discovery, and compares the results with those from the RVM. The case studies considered here are restricted to SDOF structural dynamic systems, and this is deemed sufficient to introduce and discuss the main aspects of the proposed approach. 
The layout of the paper is as follows: Section \ref{sec:model} introduces the model of the three SS prior variants used in this study, derives the MCMC procedure for sampling the model parameters, and outlines the methodology for Bayesian variable selection using SS priors. Section \ref{sec:numerical} presents numerical demonstrations of equation discovery for four SDOF oscillators that are of interest in nonlinear structural dynamics: a linear oscillator, a Duffing oscillator with cubic nonlinearity, an oscillator with quadratic viscous damping and one with Coulomb damping. Next, the proposed approach is applied to the Silverbox experimental benchmark in Section \ref{sec:exp_Silverbox}. Finally, Section \ref{sec:discussion} provides a critical discussion on the results obtained with the SS priors, and Section \ref{sec:conclusions} summarises the conclusions of the paper.

\section{Bayesian variable selection with spike-and-slab priors} \label{sec:model}
The idea of variable selection is to identify, out of $P$ basis variables in the dictionary $\mD$, the influential variables that have significant effect in explaining $\y$.
Each combination of variables corresponds to a different model, and so variable selection amounts to selecting a model from among $2^P$ possible models. In a Bayesian framework, the idea of distinguishing large effects from small effects is realised by imposing prior distributions on the weight vector $\vtheta$, such that they have a probability mass concentrated around zero and the rest over a large range of the weight space. In this sense, the SS prior -- featuring a mixture of two distributions, one with a spike at zero and the other with a diffused density over a wide range of possible values -- conforms to a conceptual ideal and is often considered as the gold standard in Bayesian variable selection \cite{polson2010shrink}. 

The first SS prior proposed for Bayesian variable selection had a spike defined by a Dirac-delta function at zero and a slab given by a uniform distribution \cite{mitchell1988bayesian}. Later on, the Dirac spike was replaced with a zero-mean Gaussian distribution with a small (but fixed) variance, and the uniform slab by another Gaussian distribution with a large variance \cite{george1993variable}. In \cite{ishwaran2005spike}, the spike and slab distributions were considered Gaussian but with bimodal priors on their variances. For a review of Bayesian variable selection strategies using SS priors, the reader is directed to \cite{ohara2009review}.

\begin{table}
	\ra{1.2}
	\centering
	\begin{tabular}{c c c c c c c}
		\toprule 
		Variant && Name && Spike distribution && Slab distribution \\ 
		\midrule 
		1 && CSS && Independent Student's-$t$ && Independent Student's-$t$\\
		2 && DSS-i && Dirac-delta && Independent Student's-$t$ \\
		3 && DSS-g && Dirac-delta && Correlated Student's-$t$ \\ 
		\bottomrule 
	\end{tabular} 
	\caption{Variants of spike-and-slab priors considered in this study.}
	\label{table:SSvariants}	
\end{table}

In this paper, three different variants of SS priors, are used for variable selection, as enumerated in Table \ref{table:SSvariants}. The first variant uses a mixture of two continuous zero-mean Student's-$t$ distributions with different (a small and a large) variances for the spike and the slab \cite{ishwaran2005spike, narisetty2014bayesian}, and is referred to as the continuous spike-and-slab (CSS) prior. The next two variants feature a mixture of a discontinuous Dirac-delta spike distribution and a continuous Student's-$t$ slab distribution, both centered at zero; they are jointly referred to as the discontinuous spike-and-slab, in short DSS, priors. The two DSS prior variants differ in their slab distributions, in that, one follows an independent Student's-$t$ and the other follows a correlated Student's-$t$ (with the correlation fashioned as Zellner's g-prior \cite{liang2008mixtures}), and are namely distinguished by their respective suffixes, DSS-i and DSS-g. The dictionary will contain many correlated variables (as will be seen later); as such, it is useful to find if a DSS prior that accounts for the correlation among the variables performs better than its independent counterpart. 

To address the two components of the SS priors, a latent indicator variable is introduced for each weight $\theta_i$. The latent indicator variable indicates the classification of a weight to one of the two components: the indicator variable takes a value one if the weight is assigned to the slab component of the prior, and zero otherwise. Since the posteriors using SS priors are analytically intractable, an MCMC-based Gibbs sampling scheme is employed to estimate the posterior probabilities of weights and indicator variables for all three SS prior variants. Variable selection is then based on the posterior probability of the indicator variable which is estimated by counting the frequency of ones. The details of the SS prior models and the procedure of posterior computation and variable selection follow next. 



\subsection{Model specification} \label{sec:SSmodel}
For basis selection with SS priors, the linear regression problem in Eq.~\eqref{eq:linreg} is considered as part of a larger hierarchical model. Treating the residual error $\err$ as a vector of i.i.d.\ Gaussian noise variables with variance $\sigma^2$, the likelihood function can be written as,
\begin{align}
	\y \mid \vtheta, \sigma^2 \sim \ND{\mD \vtheta, \sigma^2 \eye_N} 
\end{align}
where $\mathcal{N}$ stands for a Gaussian distribution.
To specify a two-component SS prior, a vector of latent indicator variables $\z = \sbk{z_1, \ldots, z_P}^T$ is introduced, where $z_i$ takes a value 0 when $\theta_i$ belongs to the spike and takes a value 1 when $\theta_i$ falls in the slab. Also, denote by $\vtheta_r \in \R^{r \times 1}$ the vector comprising those components of $\vtheta$ for which $z_i=1$. Then, the SS prior can be written as,
\begin{align}
	\pr{\vtheta \mid \z} = p_{\text{slab}}\brc{\vtheta_r} \prod_{i: z_i=0 } p_{\text{spike}}\brc{\theta_i}
\end{align}
where $p_{\text{spike}}$ and $p_{\text{slab}}$ denote the univariate spike and the multivariate slab distributions, respectively. The DSS priors considered in this study have the following forms of spike and slab distributions:
\begin{subequations} \label{eq:allvariants}
	\begin{align}
		\begin{split} \label{eq:DSS}
			\text{DSS}:& \;\; p_{\text{spike}}\brc{\theta_i} = \delta_0, \;\; \text{and} \;\; p_{\text{slab}}\brc{\vtheta_r} =  \ND{\zeros, \sigma^2 \vs \A{0,r}}, \;\;
			\A{0,r} = \begin{cases}
				\eye_r & \text{for DSS-i}\\
				N \brc{\Dr{T} \Dr{}}^{-1} & \text{for DSS-g}
			\end{cases}   
		\end{split}\\
		\begin{split}
			\text{CSS}:& \;\; p_{\text{spike}}\brc{\theta_i} = \ND{0, \sigma^2 v_1 v_{s_i}} \;\; \text{and} \;\; p_{\text{slab}}\brc{\vtheta_r} = \prod_{i:z_i=1} \ND{0, \sigma^2 v_0 v_{s_i}}  \label{eq:CSSP}
		\end{split}
	\end{align}
\end{subequations}
Note that $\mD_r$ in Eq.~\eqref{eq:DSS} is a dictionary matrix that includes only the columns of $\mD$ for which $z_i = 1$. The following points are to be noted:
\begin{itemize}
	\item The spike distributions are considered independent of the slab distributions. The spike distribution in DSS is modelled by a Dirac-delta function at zero, denoted by $\delta_0$, whereas that in CSS is modelled by a zero-mean small-variance continuous distribution.
	\item The slab distributions for all three variants have their mean centred at zero and their (co-)variances proportional to the product of measurement noise variance $\sigma^2$ and slab variance $\vs$. The inclusion of the measurement noise in the prior allows it to scale naturally with the scale (i.e.\ the measurement units) of the outcome $\y$.
	\item It is natural to assume weight-specific slab variances that can modify the strength of each individual prior \cite{narisetty2014bayesian}. While the CSS priors feature weight-specific slab variances $v_{s_i}$, a common slab variance $\vs$ is assumed for all weights in the DSS priors. A common slab variance is found to yield better results for DSS priors.
	\item The difference in the variances of the spike and slab distributions in the CSS case is facilitated by the use of constants $v_0$ and $v_1$ such that $v_0 \ll v_1$, leading to a narrow spike and a wide slab.
	\item The covariance of the slab distribution of the DSS-g prior includes an additional scaling by the Fischer information matrix $N \brc{\Dr{T} \Dr{}}^{-1}$, which accounts for the correlation among the basis variables; in contrast, the DSS-i prior uses an independent slab distribution over each component of $\vtheta$.
\end{itemize}
The marginal Student's-$t$ distributions for the respective slabs (and spikes in case of CSS priors) given the noise variance are achieved by imposing an inverse-Gamma prior on the slab variance $\vs$ (or equivalently on each $v_{s_i}$ for CSS priors),  
\begin{align}\label{eq:vs_prior}
	\vs \sim \IG{a_v, b_v} 
\end{align}
The transformation of the Gaussian to Student's-$t$ prior on $\vtheta$ via the inverse-Gamma prior on $\vs$ is because of the scale mixture property of Gaussians \cite{andrews1974scale}. One could have alternatively used an exponential prior on $\vs$ to obtain a marginal Laplace prior on $\vtheta$, or simply treated $\vs$ as a constant to impose a Gaussian prior. The motivation for modelling the slab using Student's-$t$ distributions lies in being able to provide a fair comparison with the RVM, which also uses a Student's-$t$ prior. 

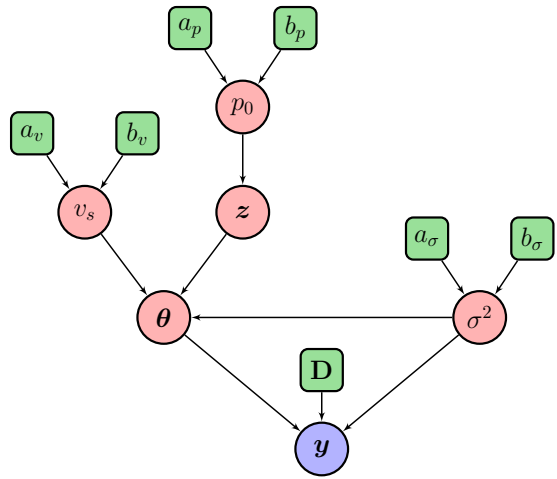
\begin{figure}[h]
	\begin{center}
		\scalebox{.7}{
\tikzstyle{rv} = [circle,  align=center, very thick, text centered, draw=black, fill=red!30, minimum size=1cm,font = \Large]
\tikzstyle{dv} = [rectangle,  align=center,very thick, rounded corners, minimum width=0.8cm, minimum height=0.8cm,text centered, draw=black, fill=green!70!black!40!, font = \Large\sffamily]
\tikzstyle{arrow} = [thick,-latex',>=stealth]
\tikzstyle{line} = [draw, -latex']

\begin{tikzpicture} [node distance = 2cm]
		\node (y) [rv, fill=blue!30] {$\ve{y}$};
		\node (D) [dv, above of=y, yshift = -0.5cm] {$\matr{D}$};
		\node (sig) [rv, above of=y, xshift = 3cm, yshift = 0.5cm] {$\sigma^2$};
		\node (sa) [dv, above of=sig, xshift = -1cm, yshift = -0.5cm] {$a_{\sigma}$};
		\node (sb) [dv, above of=sig, xshift = 1cm, yshift = -0.5cm] {$b_{\sigma}$};
		
		\node (theta) [rv, above of=y, xshift = -3cm, yshift = 0.5cm] {$\ve{\theta}$};
		\node (vs) [rv, above of=theta, xshift = -1.5cm] {$v_s$};
		\node (vsa) [dv, above of=vs, xshift = -1cm, yshift = -0.5cm] {$a_v$};
		\node (vsb) [dv, above of=vs, xshift = 1cm, yshift = -0.5cm] {$b_v$};
		\node (z) [rv, above of=theta, xshift = 1.5cm] {$\ve{z}$};
		\node (p0) [rv, above of=z] {$p_0$};
		\node (p0a) [dv, above of=p0, xshift = -1cm, yshift = -0.5cm] {$a_p$};
		\node (p0b) [dv, above of=p0, xshift = 1cm, , yshift = -0.5cm] {$b_p$};
		
		\draw [arrow] (D) -- node {} (y);
		\draw [arrow] (sig) -- node {} (theta);
		
		\draw [arrow] (sig) -- node {} (y);
		\draw [arrow] (sa) -- node {} (sig);
		\draw [arrow] (sb) -- node {} (sig);
		
		\draw [arrow] (theta) -- node {} (y);
		
		\draw [arrow] (vs) -- node {} (theta);
		\draw [arrow] (vsa) -- node {} (vs);
		\draw [arrow] (vsb) -- node {} (vs);

		\draw [arrow] (z) -- node {} (theta);
		
		\draw [arrow] (p0) -- node {} (z);
		\draw [arrow] (p0a) -- node {} (p0);
		\draw [arrow] (p0b) -- node {} (p0);

\end{tikzpicture}}
		\caption{Graphical structure of the hierarchical spike-and-slab model for linear regression; the variables in circles represent random variables, while those in squares represent deterministic parameters. Note, in case of DSS priors, the slab variance $\vs$ is a scalar, while that for CSS priors would be a vector of weight-specific slab variances.}
		\label{f:PGM_SNS}
	\end{center}
\end{figure}

In SS priors, each latent indicator variable $z_i$ is assigned an independent Bernoulli prior, controlled by a common hyperparameter $\pz$,
\begin{align} \label{eq:bern}
	z_i \mid \pz \sim \text{Bern}(\pz)
\end{align} 
Eq.~\eqref{eq:bern} implies that the selection of a basis variable from the dictionary $\mD$ is independent of the inclusion of any other basis variables in $\mD$. The hyperparameter $\pz$ in Eq.~\eqref{eq:bern} represents the fraction of the total basis variables in $\mD$ that are \textit{a priori} expected to be selected in the final model; it can be assigned a fixed value. For example, $\pz = \frac{1}{2}$ implies that each basis variable in $\mD$ has equal chance of being selected and reflects the prior belief that the model should include approximately half of the basis variables in $\mD$. However, here $\pz$ is allowed to be adaptively refined by the data via a Beta prior,
\begin{align} \label{eq:p0_prior}
	\pz \sim \text{Beta}(a_p, b_p)
\end{align}
Finally, the measurement noise variance $\sigma^2$ is assigned an inverse-Gamma prior,
\begin{align} \label{eq:sigma_prior}
	\sigma^2 \sim \IG{a_{\sigma}, b_{\sigma}} 
\end{align} 
Note that $a_v$, $b_v$, $a_p$, $b_p$, $a_{\sigma}$, $b_{\sigma}$ appearing in Eqs.~\eqref{eq:vs_prior}, \eqref{eq:p0_prior} and \eqref{eq:sigma_prior} are deterministic hyperparameters, controlling the shape of the respective hyper-priors. The complete hierarchical SS model for linear regression is illustrated in Figure~\ref{f:PGM_SNS}.

\subsection{Posterior computation} \label{sec:Gibbs_samp}
Once the hierarchical form of the SS priors is specified, the next part entails extracting the information relevant to variable selection from the posteriors of $\z$, $\vtheta$ and $\sigma^2$. The joint posterior of $\pr{\z, \vtheta, \sigma^2 \mid \y}$ can be computed using Bayes' theorem in the form,
\begin{align} \label{eq:joint_post_dist}
	\pr{\z, \vtheta, \sigma^2 \mid \y} = \frac{\pr{\y \mid \vtheta, \sigma^2} \pr{\vtheta, \z} \pr{\sigma^2}}{\pr{\y}}
\end{align} 
where $\pr{\y \mid \vtheta, \sigma^2}$ is the likelihood, $\pr{\vtheta, \z}$ is the joint prior over weights and latent indicator variables, $\pr{\sigma^2}$ is the prior over measurement noise, and $\pr{\y}$ is the normalising constant. 
Exact Bayesian inference is difficult with SS priors, and often MCMC techniques are employed to sample from the posteriors \cite{geweke1996variable,malsiner2011comparing}. In this case, a Gibbs sampler \cite{casella1992explaining} is used to draw samples from the posterior. 
Gibbs sampling needs knowledge of the full conditional distributions which can be derived analytically with the use of conjugate priors. It should be mentioned that the sampling schemes for the DSS and CSS priors differ slightly as a result of the need to integrate out the Dirac-delta function in the case of DSS priors. The Gibbs sampling scheme for the CSS prior has been adopted from \cite{narisetty2014bayesian}, and details of the sampling steps are provided in \ref{sec:MCMCsteps_CSS}. Below, the Gibbs sampling steps for the parameters $\vtheta$, $\z$, $\pz$, $\vs$ and $\sigma^2$ of the DSS priors are provided. 
\begin{enumerate} [label=(\alph*)]
	\item The components of $\vtheta$ that correspond to $z_i=0$ (i.e.\ belong to the Dirac-delta spike) are set to zero. The rest of the  components belonging to the slab, represented by $\vtheta_{r}^{}$, are sampled as follows,
	\begin{align} \label{eq:sampleW}
		\vtheta_{r}^{} \mid \y, \z, \vs, \sigma^2 \sim \ND{\muz^{}, \sigma^2 \Sigmaz^{}}
	\end{align}
	where $\Sigmaz^{} = \brc{\Dr{T} \Dr{} + \vs^{-1} \A{0,r}^{-1}}^{-1}$ and $\muz^{} = \Sigmaz^{} \Dr{T} \y$.
	\item $\sigma^2$ is sampled from an inverse Gamma distribution as follows,
	\begin{align} \label{eq:sampleSig}
		\sigma^2 \mid \y, \z, \vs \sim \IG{a_{\sigma} + \frac{N}{2}, \; b_{\sigma} + \frac{1}{2} \brc{\y^T \y - \muz^T \Sigmaz^{-1} \muz^{}} }
	\end{align}
	\item $\vs$ is sampled from an inverse Gamma distribution as follows,
	\begin{align} \label{eq:sampleVs}
		\vs \mid \vtheta, \z, \sigma^2 \sim \IG{a_v + \frac{s_z}{2}, \; b_v + \frac{1}{2 \sigma^2} \vtheta_r^T \A{0,r}^{-1} \vtheta_r^{} }
	\end{align}
	where $s_z = \sum_{i=1}^{P} z_i$.
	\item $\pz$ is sampled from a Beta distribution as follows,
	\begin{align} \label{eq:sampleP}
		\pz \mid \z \sim \text{Beta} \brc{a_p + s_z, \; b_p + P - s_z}
	\end{align}
	\item The conditional distribution of $\z$ is expressed componentwise. The odds of $z_i= 1$ to $z_i = 0$ are computed, given the values of other $\z$ components, denoted here as $\z_{-i}$. The components of $\z$ are sampled (in a random order) as follows, 
	\begin{align} \label{eq:sampleZ}
		z_i \mid \y, \vs, \pz &\sim \text{Bern}(\xi_i),  \text{   with } \xi_i = \frac{\pz}{\pz + \frac{\pr{\y \mid z_i = 0, \z_{-i}, \vs} }{\pr{\y \mid z_i = 1, \z_{-i}, \vs} }  (1-\pz)}
	\end{align}
	In the above sampling step, the marginal likelihood $\pr{\y \mid \z, \vs}$  after integrating out $\vtheta$ is computed using,
	\begin{align} \label{eq:add_step_DSS}
		\pr{\y \mid \z, \vs} &= \frac{\Gamma({a}_{\sigma} + 0.5N)}{(2\pi)^{N/2} (\vs)^{s_z/2}} 
		\frac{\brc{b_{\sigma}}^{a_{\sigma}}}{\Gamma(a_{\sigma})}
		\frac{\brc{\det \brc{\A{0,r}^{-1}}}^{1/2} \;\; \brc{\det \brc{\brc{\Dr{T} \Dr{} + \vs ^{-1} \A{0,r}^{-1}}^{-1} }}^{1/2}}{\brc{{b}_{\sigma} + 0.5 \y^T \brc{\eye_N -  \Dr{} \brc{\Dr{T} \Dr{} + \vs ^{-1} \A{0,r}^{-1}}^{-1} \Dr{T} } \y}^{\brc{{a}_{\sigma} + 0.5N}}}
	\end{align}
	where $\Gamma(\cdot)$ denotes the Gamma function and $\det(\cdot)$ denotes the determinant operator.
\end{enumerate}

\noindent By repeated successive sampling using Eqs.~\eqref{eq:sampleW} to \eqref{eq:sampleZ}, the following Markov chain is produced,
\begin{align}
	\vtheta^{(0)}, {\sigma^2}^{(0)}, \vs^{(0)}, \pz^{(0)}, \z^{(0)}, \ldots, \vtheta^{(l)}, {\sigma^2}^{(l)}, \vs^{(l)}, \pz^{(l)}, \z^{(l)}, \ldots
\end{align}
which embeds the Markov chains for $\z$, $\vtheta$ and $\sigma^2$. The first few samples of the chain are discarded as burn-in, and the remaining $J$ samples are used for basis variable selection, as described next.


\subsection{Basis selection and posterior prediction} \label{sec:basis_sel}
As mentioned previously, there are $2^P$ models possible with $P$ basis variables in the dictionary, where a model is indexed by which of the $z_i$s equal one and which equal zero. For example, the model with zero basis variables has $\z=\zeros$, whereas the model that includes all basis variables has $\z=\ones$. Finding the model with highest posterior probability is often challenging when $P$ is large, as one would probably need more than $2^P$ samples to explore the entire space of models.
In this work, the marginal posterior inclusion probabilities (PIP), $\pr{z_i = 1\mid \y}$, are used to select those basis variables whose corresponding probability is more than a fixed probability threshold. The PIPs are approximated using $J$ Gibbs samples, as follows,
\begin{align}
	\pr{z_i = 1\mid \y} \approx \frac{1}{J} \sum_{j=1}^{J}  \mathbb{I}\brc{z_i^{(j)}= 1}
\end{align}
where $\mathbb{I}\brc{\cdot}$ stands for an indicator function.
Specifically, one selects the $i^{\text{th}}$ basis variable from $\mD$ and includes it in the final model if,
\begin{align} \label{eq:selection_criterion}
	\pr{z_i =1 \mid \y} > 0.5
\end{align}
The higher the posterior mean of the indicator variable, the higher is evidence that the parameter $\theta_i$ might be different from zero and therefore the corresponding basis variable will have an impact on $\y$. The above criterion implies the selection of basis variables that appear in at least half of the visited models. The final estimated model $\hat{\F}$ so obtained corresponds to the median probability model \cite{barbieri2004optimal}, and is computationally advantageous since estimating this model often requires fewer Gibbs iterations than are required for the highest probability model. The model $\hat{\F}$ would correspond to the discovered governing equations of motion in this study. Post variable selection, the estimated mean and covariance of the parameter vector $\vtheta$, denoted by $\hat{\muz}_{\vtheta}$ and $\hat{\Sigmaz}_{\vtheta}$, respectively, will feature non-zero components only at indices corresponding to those of the selected basis variables.   

Subsequently, predictions with the estimated model $\hat{\F}$ can be performed using $\hat{\muz}_{\vtheta}$ and $\hat{\Sigmaz}_{\vtheta}$ via the expressions,
\begin{align}
	\ve{\mu}_{\y^{*}} &= \mD^{*} \hat{\ve{\mu}}_{\vtheta} \\
	\matr{\Sigma}_{\y^{*}} &= \mD^{*} \hat{\matr{\Sigma}}_{\vtheta} {\mD^{*}}^T + \hat{\mu}_{\sigma^2} \eye_{N^*}
\end{align}
where $\mD^{*} \in \R^{N^* \times P}$ is the test dictionary, defined at a set of $N^*$ previously-unseen test data points, $\ve{\mu}_{\y^{*}} \in \R^{N^* \times 1}$ is the predicted mean of the target test vector, $\matr{\Sigma}_{\y^{*}} \in \R^{N^* \times N^*}$ is the predicted covariance of the target test vector, and $\hat{\mu}_{\sigma^2} \in \R$ is the mean of the measurement noise variance estimated using $J$ Gibbs samples of $\sigma^2$.

\section{Numerical studies} \label{sec:numerical}
In this section, the performance of the proposed sparse Bayesian algorithms in discovering governing equations is investigated. SDOF oscillators of the form expressed by Eq.~\eqref{eq:main_eqn} containing the nonlinear term $g(x_1, x_2)$ are considered, where $x_1$ and $x_2$ represent the displacement and velocity states of the oscillator. Different forms of the nonlinearity $g(x_1, x_2)$ lead to different systems of engineering interest. Four different cases of nonlinearities $g(x_1,x_2)$ are considered in this study, as enumerated in Table~\ref{table:nonl_types}.

\begin{table}[ht]
	\centering
	\begin{tabular}{cclclcll}
		\toprule[1.5pt]
		\textbf{System} && \textbf{Name} && $g(x_1,x_2)$ &&  \\ \midrule
		1 && Linear &&  0      &&   \\
		2 && Duffing && $k_3 x_1^3$ && $k_3 = 10^5$ \\
		3 && Quadratic viscous damping && $c_2 {x}_2 |{x}_2|$ && $c_2 = 2$ \\ 
		4 && Coulomb friction damping && $c_F \text{sgn}\brc{x_2}$ && $ c_F = 1$ \\
		\bottomrule                        
	\end{tabular}
	\caption{Simulation cases.}
	\label{table:nonl_types}	
\end{table}

The first system is a linear system, used here to verify if the proposed method is capable of ruling out the existence of any nonlinearities in the dynamical system. The second system is a Duffing oscillator, with a cubic displacement nonlinearity $g(x_1,x_2) = k_3 x_1^3$; it can be used to represent many physical systems and has been widely used in a large number of studies in nonlinear system identification \cite{kerschen2006past}. In structural systems, the nonlinearity can be used to represent hardening geometric nonlinearity arising as a result of large displacements; as the displacement increases, the nonlinear restoring force becomes greater than that expected from the linear term alone. The third system includes a quadratic viscous damping nonlinearity $g(x_1,x_2) = c_2 {x}_2 |{x}_2|$, where $|\cdot|$ denotes the absolute value. This type of damping occurs in fluid flows through orifices or around a slender member. The former situation is common in automotive dampers, whereas the latter occurs in fluid loading of offshore structures \cite{worden2019nonlinearity}. The fourth system includes a Coulomb friction damping nonlinearity $g(x_1,x_2) = c_F \text{sgn}\brc{{x}_2}$, where $\text{sgn}\brc{\cdot}$ denotes the signum function. This type of nonlinearity is encountered in situations that involve interfacial motion or sliding \cite{dahl1976solid}, such as dry sliding occurring in bolted joints. The four SDOF systems are simulated using the following parameters:
\begin{itemize}
	\item The parameters of the linear system are taken as: $m = 1$, $c = 2$, and $k = 1000$.
	\item The three other nonlinear systems use the same values of parameters for the underlying linear part and only differ in the additional nonlinear term $g(x_1,x_2)$. The respective forms and the values of $g(x_1,x_2)$ are provided in Table~\ref{table:nonl_types}.
	\item The systems are excited using a bandlimited -- passband $[0,100]\si{Hz}$ -- Gaussian excitation with zero mean and standard deviation of $50$.
	\item The displacement $x_1$ and velocity $x_2$ for each system are simulated using a fixed-step fourth-order Runge-Kutta numerical integration scheme, with a sampling rate of $1000$Hz.
	\item The acceleration $\dot{x}_2$ is obtained using Eq.~\eqref{eq:main_eqn}.
\end{itemize}


Before commencing equation discovery, one requires the knowledge of the time-series data of displacement, velocity, acceleration and input force signals from forced vibration testing of the system; the acceleration data is used as the measurement vector $\y$ whereas the displacement, velocity, and input force data are used in composing the basis variables of the dictionary $\mD$ (see Eq.~\eqref{eq:linreg_expand}). 
It is assumed that noisy measurements of all input and outputs, i.e., displacement $x_1$, velocity $x_2$, acceleration $\dot{x}_2$, and input force $u$ are available, and the noisy signals are used to compose the dictionary $\mD$ and the target measurement vector $\y$. The noise in the measurements is modelled as sequences of zero-mean Gaussian white noise with a standard deviation equal to 5\% of the standard deviation of the simulated quantities.

In this work, the dictionary $\mD$ is constructed with 36 basis variables, where each basis variable represents a certain function of the states $x_1$ and $x_2$: 
\begin{align} \label{eq:dictionary}
	\mD = \cbk{P^1(\x), \ldots, P^{6}(\x), \text{sgn}\brc{\x}, |\x|, \x \otimes |\x|, u}
\end{align} 
Here, $P^{\gamma}(\x)$ denotes the set of terms in the polynomial expansion of the sum of state vectors $(x_1 + x_2)^{\gamma}$. The dictionary consists of basis variables that are terms from polynomial orders up to $\gamma = 6$ and certain other terms. The term $\text{sgn}\brc{\x}$ represents the signum functions of states, i.e., $\text{sgn}\brc{x_1}$ and $\text{sgn}\brc{x_2}$. Similarly, $|\x|$ denotes the absolute functions of states, i.e., $|x_1|$ and $|x_2|$. The tensor product term $\x \otimes |\x|$ represents the set of functions: $x_1 |x_1|$, $x_1 |x_2|$, $x_2 |x_1|$ and $x_2 |x_2|$. Note that the total number of models that can be formed by combinatorial selection of all 36 basis variables in the dictionary is $2^{36}$, and grows exponentially as the number of basis variables increases. 

An issue with the constructed dictionary in Eq.~\eqref{eq:dictionary} is that it is often ill-conditioned. This happens due to a combined effect of (a) the large scale difference among the basis variables and (b) the presence of strong linear correlation between certain basis variables. Appropriate scaling of the columns can help to reduce the difference in scales and improve the conditioning of the dictionary. For the purpose of Bayesian inference, the columns of the training dictionary are normalised (i.e.\ they are centered and scaled to have zero mean and unit standard deviation). Additionally, the training measurement data are detrended to have zero mean; as such there is no need to include a constant intercept term in the dictionary. Put formally, the training dictionary and the target vector $\brc{\mD_s, \y_s}$ input to the Bayesian inference algorithm have the forms, 
\begin{align}
	\begin{split}
		\mD^s &= \brc{\mD - \ve{1} \ve{\mu}_{\mD}} \ve{S}^{-1}_{\mD}  \\
		\y^s &= \y - \ve{1} \mu_{\y}
	\end{split}
\end{align}
where $\ve{1}$ denotes a column vector of ones, $\ve{\mu}_{\mD}$ is a row vector of the column-wise means of $\mD$, $\ve{S}_{\mD}$ is a diagonal matrix of the column-wise standard deviations of $\mD$, and $\mu_{\y}$ is the mean of the training target measurement vector $\y$. Note that this modification implies that, post Bayesian inference, the estimated mean and covariance of the scaled coefficients $\vtheta^s$, denoted by $\hat{\ve{\mu}}_{\vtheta^s}$ and $\hat{\matr{\Sigma}}_{\vtheta^s}$, have to be transformed back to the original space using the relations,
\begin{align}
	\begin{split}
		\hat{\ve{\mu}}_{\vtheta} &= \ve{S}^{-1}_{\mD} \hat{\ve{\mu}}_{\vtheta^s} \\
		\hat{\matr{\Sigma}}_{\vtheta} &= \ve{S}^{-1}_{\mD} \hat{\matr{\Sigma}}_{\vtheta^s} \ve{S}^{-1}_{\mD}
	\end{split}
\end{align} 
%
%

For Bayesian inference with the SS priors, the Gibbs sampler is commenced with the following initial values of the hyperparameters: 
$\pz^{(0)} = 0.1$, $\vs^{(0)} = 10$, and ${\sigma^2}^{(0)}$ is set equal to the residual variance from ordinary least-squares regression. Additionally, for the CSS prior, each component of the vector of weight-specific slab variances is initialised to the value of $\vs^{(0)}$, and the two variance-scaling constants are set to $v_0 = \frac{1}{N}$ and $v_1 = 100 v_0$, respectively. Note that both $v_0$ and $v_1$ are made to depend on the sample size to ensure model selection consistency \cite{narisetty2014bayesian}. To facilitate faster convergence of the Gibbs sampler to a good solution, the initial vector of binary latent variables $\z^{(0)}$ is computed by starting off with $z_1, \ldots, z_P$ set to zero and then activating the components of $\z$ that reduce the mean-squared error on the (training) data, until an integer number $(\approx \pz^{(0)} P)$ of components of $\z$ are equal to one. 
Given all the other parameters, the initial value of $\vtheta^{(0)}$ is obtained by sampling from Eq.~\eqref{eq:sampleW}. The deterministic prior parameters are set to the following values: $a_p=0.1$, $b_p=1$ are chosen for the Beta prior on $\pz$ to promote selection of sparse models, $a_v=0.5$, $b_v=0.5$ for inverse-Gamma prior on slab variance, 
and $a_{\sigma} = 10^{-4}$, $b_{\sigma}=10^{-4}$ are chosen for a non-informative prior on measurement noise. Four Markov chains are used for Gibbs sampling with 5000 samples in each chain. The first 1000 samples of each chain are discarded as burn-in, and the remaining $4000 \times 4 = 16000$ samples are used for posterior computation. To ensure variability across the chains, each of them is initialised with randomly perturbed values of the aforementioned initial hyperparameters. The multivariate potential scale reduction factor $\hat{R}$ \cite{brooks1998general}, which estimates the potential decrease in the between-chain variance with respect to the within-chain variance, is applied to assess the convergence of the generated samples of $\vtheta$; a value of $\hat{R} < 1.1$ is adopted to decide if convergence has been reached. 

\begin{figure}[H]
	\begin{center}
		\includegraphics[scale=0.8]{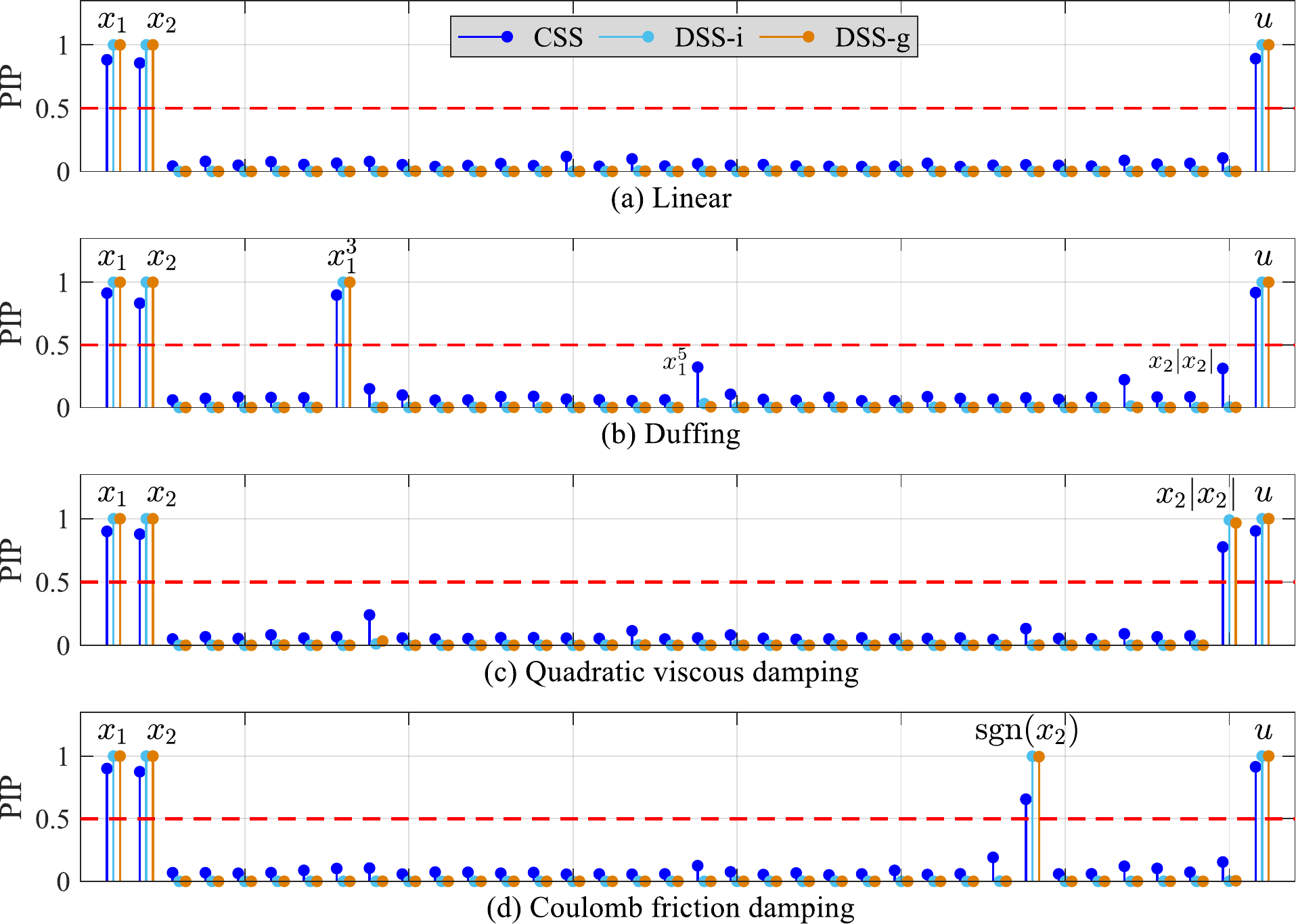}
		\caption{Basis variable selection based on marginal posterior inclusion probability (PIP), $\pr{z_i=1 \mid \y}$. The horizontal axes represent the collection of 36 basis variables; the variables having marginal PIP > 0.5 are included in the final estimated model.}
		\label{f:SS_varsel}
	\end{center}
\end{figure}

Figure~\ref{f:SS_varsel} demonstrates the procedure of variable selection for the four systems, based on the marginal PIP, $\pr{z_i=1 \mid \y}$,  $i = 1,\ldots,36$. When $\pr{z_i=1 \mid \y}=1$, it implies that the $i^{\text{th}}$ basis variable had been selected in all Gibbs posterior samples, while $\pr{z_i=1 \mid \y} = 0$ implies the $i^{\text{th}}$ basis variable has never been selected. As mentioned in Section \ref{sec:basis_sel}, only those basis variables are included in the final estimated model whose corresponding marginal PIPs are greater than the set threshold of 0.5 (shown by the dotted line in red in Figure~\ref{f:SS_varsel}). It can be seen that the estimated models for all the four systems are able to select the true basis variables out of the pool of 36 basis variables. For DSS priors, the computed marginal PIPs corresponding to the true basis variables are close to one, which indicates a strong selection probability. However, the selection probabilities with CSS priors are not as strong; they exhibit smaller PIPs for the relevant variables, compared to those from DSS priors. The weaker selection probability with CSS priors is apparent in the Duffing oscillator case, where the true relevant variables $x_1$ and $x_1^3$ draw marginal PIPs of around 0.9, while a irrelevant variable $x_1^5$ receives a marginal PIP of 0.3. Similarly, $x_2$ is selected with PIP of 0.8 whereas $x_2 \abs{x_2}$ gets discarded with a PIP of 0.3. Although this behaviour occurs more frequently with CSS priors, it can also happen with DSS priors, especially in situations when there are strong correlations between certain basis variables causing the Bayesian algorithm to be confused as to which of the set of correlated basis variables should be selected.

\begin{figure}[H]
	\begin{center}
		\includegraphics[scale=0.8]{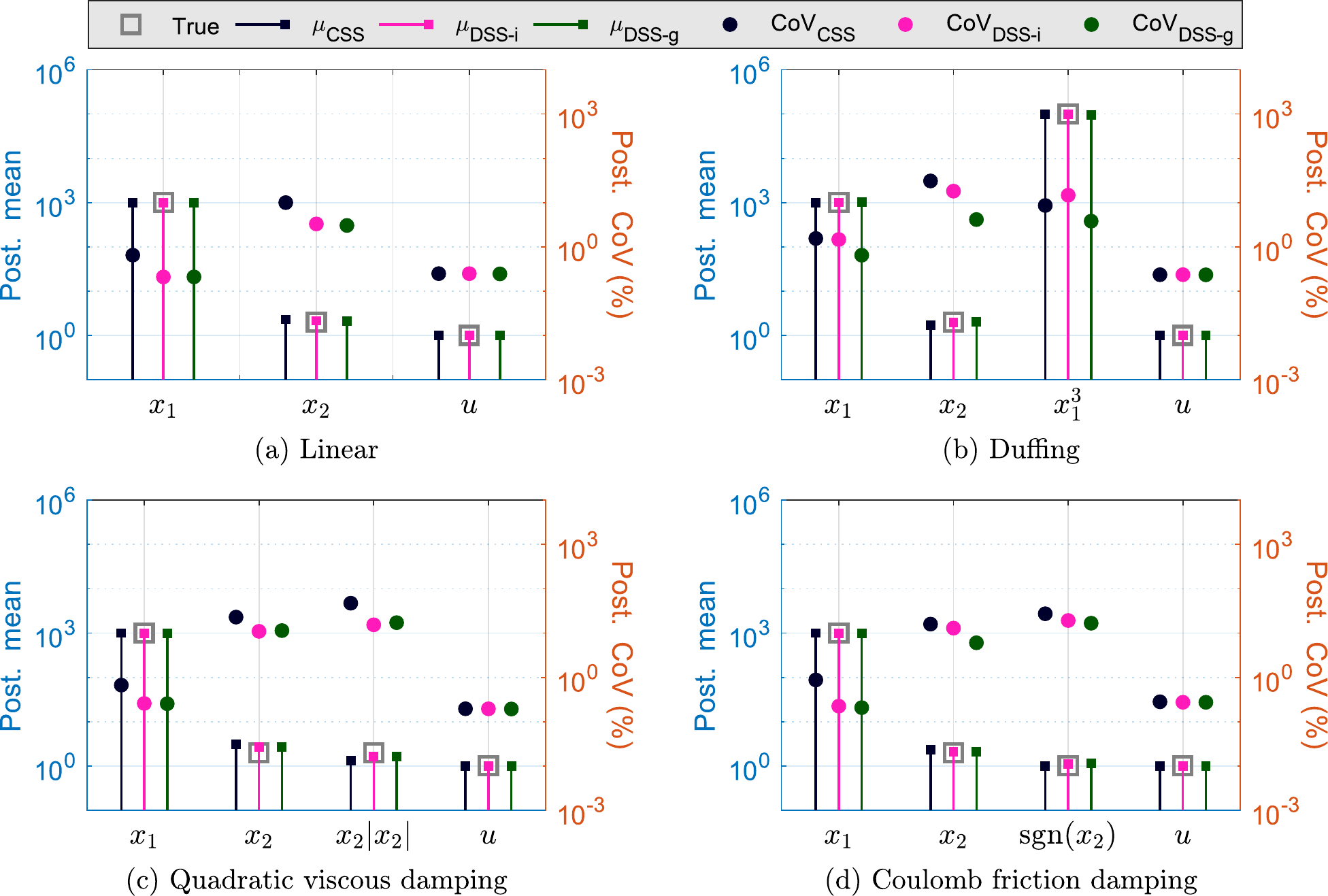}
		\caption{Estimates of parameters of the selected basis variables. The left vertical axes represents absolute posterior means and the right axes illustrated the associated coefficient of variations (CoVs) of the estimated parameters inferred using CSS, DSS-i and DSS-g priors.}
		\label{f:SS_PE}
	\end{center}
\end{figure}

\begin{figure}[h!]
	\begin{center}
		\includegraphics[scale=0.75]{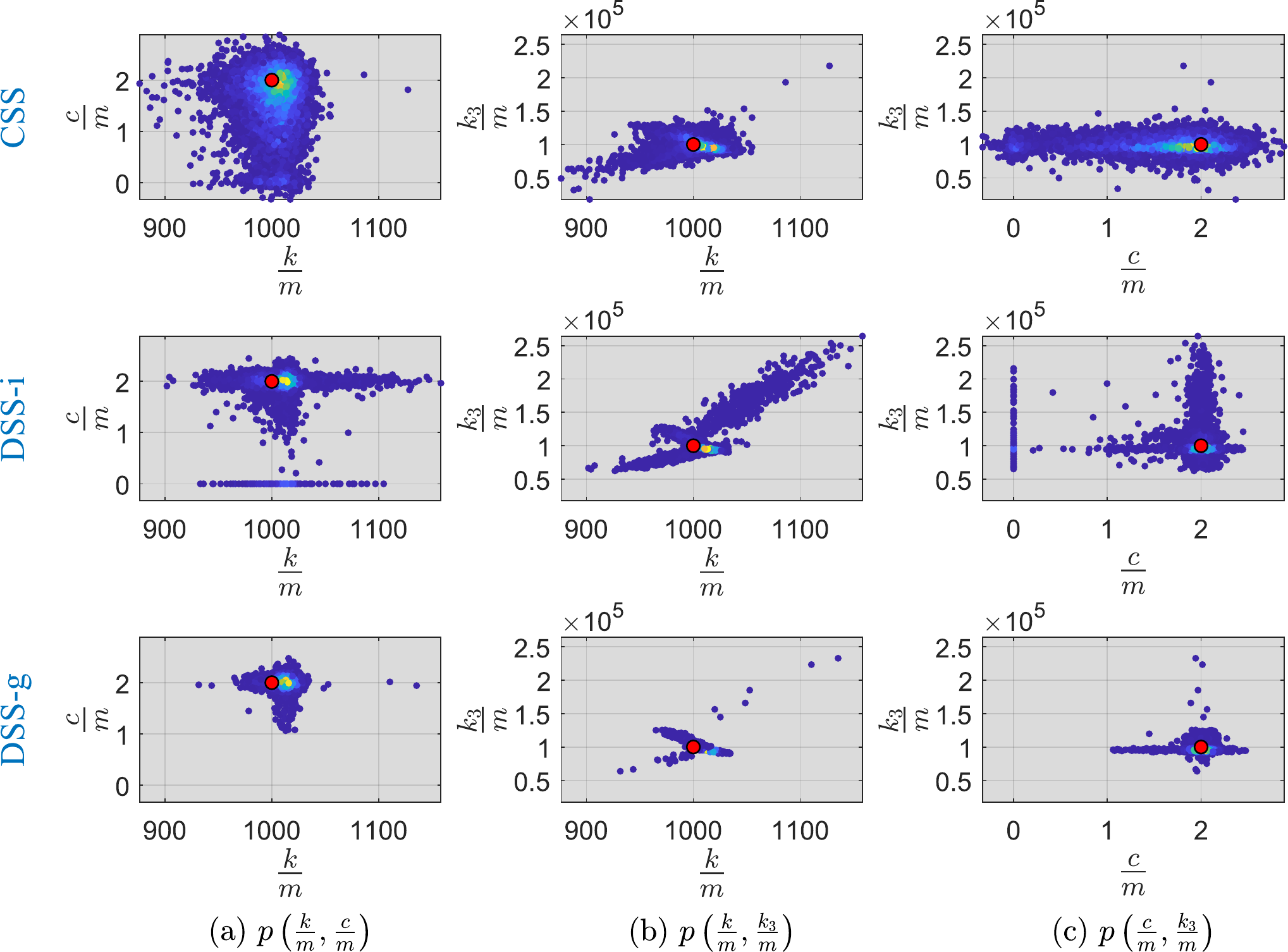}
		\caption{Pairwise joint posteriors of the parameters $\frac{k}{m}$, $\frac{c}{m}$, and $\frac{k_3}{m}$ corresponding to the selected basis variables $x_1$, $x_2$, and $x_1^3$, for the Duffing oscillator case. The red circles indicate the true values of parameters.}
		\label{f:SS_Duffing_densities}
	\end{center}
\end{figure}

Figure~\ref{f:SS_PE} plots the absolute posterior means and coefficient of variations (CoVs) of the parameters that correspond to the selected basis variables in Figure~\ref{f:SS_varsel}; the CoVs are expressed as percentage ratios of the means to the standard deviations. The mean values of the parameters are found to agree very well with the corresponding true values, and the posterior CoVs are quite small for the parameters associated with variables $u$ and $x_1$. Higher COVs are seen for parameter estimates associated with variables that are functions of $x_2$. The greater uncertainty associated with variables $x_2$ and functions of $x_2$ is because of two factors: (a) small values of the parameters associated with these variables and (b) presence of many other correlated variables in the dictionary, which, coupled together, confuses the Bayesian learner. It is also noted that the posterior standard deviations of the parameters inferred with CSS priors are comparatively larger than those inferred with the DSS priors. For illustration, the pairwise joint posteriors of the parameters for the Duffing oscillator case are plotted in Figure~\ref{f:SS_Duffing_densities}. Clearly, the posterior samples with CSS priors can be seen to spread over a larger parameter space in the plots. This behavior with CSS priors is caused by the less restrictive continuous spike distribution, which occasionally allows the irrelevant variables to take non-zero weights and biases the weights of the relevant parameters, thereby inducing a greater spread in the weights (or parameters) of the relevant variables. The posterior mean values of the parameters inferred with CSS priors are, however, found to agree well with the true values of the parameters. It should also be mentioned that the posteriors obtained with SS priors can be multi-modal, as seen from Figure~\ref{f:SS_Duffing_densities}, . 

\subsection*{Equation discovery performance comparison using Monte Carlo simulations} \label{sec:MonteCarlo}
In this section, Monte Carlo simulations are used to assess the equation discovery performance of the proposed MCMC algorithms with SS priors -- hereafter referred to as the MCMC-SS algorithms. Furthermore, the popular RVM algorithm \cite{tipping2001sparse, tipping2003fast} using a Student's-$t$ prior is implemented for the sake of comparison of equation discovery results. The freely available \textit{SparseBayes} software \cite{RVMsoftware} is used for implementing the RVM.

%
%
1000 different realisations for each of the four systems, as summarised in Table~\ref{table:nonl_types}, were considered. The realisations were created by introducing random perturbations of $0.1 \kappa$ to the nominal values of the parameters $c, k, k_3, c_2, c_F$, such that the new realisations have parameters $\bar{c} = (1 + 0.1 \kappa) c$, \;$\bar{k} = (1 + 0.1 \kappa)  k$, and so on. The variable $\kappa$ was sampled from a standard Gaussian distribution $\ND{0,1}$ for each realisation. Note that the nominal values of parameters are the ones that were used in the previous numerical study. In order to assess the performance, the following performance metrics are defined: \begin{itemize}
	\item Weight estimation error, $e_{\theta}= \frac{\norm{\hat{\vtheta} - \vtheta}_2}{\norm{\vtheta}_2} $, where $\hat{\vtheta}$ is the estimate of the true weight vector $\vtheta$. In the case of SS priors, $\hat{\vtheta}$ is obtained as the mean estimate of the posterior sample, whereas in the case of RVM, it is obtained as the maximum \textit{a posteriori} estimate. Similarly, one can also define a scaled weight estimation error, $e_{\theta^s}= \frac{\norm{\ve{S}_{\mD}\brc{\hat{\vtheta} - \vtheta}}_2}{\norm{\ve{S}_{\mD} \vtheta}_2}$. 
	\item Test set prediction error, $e_{p}= \frac{\norm{\y^* - \mD^* \hat{\vtheta}}_2}{\norm{\y^*}_2}\times 100$, where $\y^*$ is the test set of responses, $\mD^*$ is the unscaled test dictionary, and $\hat{\vtheta}$ is the estimate of the unscaled weight vector obtained using training data. 2000 data points were used for training and another 2000 data points for testing.
	\item False discovery rate (FDR), defined as the ratio of the number of false basis variables selected to the total number of basis variables selected in the estimated model. A good variable selection algorithm should output a model with fewer false discoveries and result in a low FDR.
	\item Exact model selection indicator, denoted by $\hat{\mathcal{M}}=\mathcal{M}$, is an indicator variable that takes value 1 when the estimated model $\hat{\mathcal{M}}$ has the exact same basis variables as the true model $\mathcal{M}$, and is zero otherwise. 
	\item Superset model selection indicator, denoted by $\hat{\mathcal{M}} \supset \mathcal{M}$, is an indicator variable that takes value 1 when the estimated model $\hat{\mathcal{M}}$ includes all the basis variables present in the true model $\mathcal{M}$, and is zero otherwise. 
\end{itemize}

The above performance metrics are evaluated for each of the 1000 different realisations for all four systems, and the averages of the results are reported in Table \ref{table:compareMCS_A}.

\begin{table}[h!] 
	\centering
	\ra{0.6}
	\begin{tabular}{ccccccccc}
		\toprule[1.5pt]
		Type & {Alg.} & 	$e_{\theta^s}$ &	$e_{\theta}$ & $e_p$ & FDR & $\hat{\mathcal{M}}=\mathcal{M}$ & $\hat{\mathcal{M}} \supset \mathcal{M}$\\ \midrule
		%
		%
		\multirow{4}{*}{Linear} 
		%
		& \textcolor{green!50!black!90!}{RVM} 
		& 0.010 & 502.261 & 0.097 & 0.576 & 0.005 & {0.999} \\
		& \textcolor{red}{MCMC-CSS} 
		& {0.006} & 6.735 & 0.074 & {0.002} & {0.993} & {0.995} \\
		& \textcolor{violet}{MCMC-DSS-i} 
		& 0.004 & 6.479 & {0.073} & {0.002} & {0.993} & {1.000} \\
		& \textcolor{blue}{MCMC-DSS-g} 
		& \textbf{0.004} & \textbf{1.513} & \textbf{0.071} & \textbf{0.001} & \textbf{0.997} & \textbf{1.000} \\
		%
		%
		\midrule
		\multirow{4}{*}{Duffing} 
		%
		& \textcolor{green!50!black!90!}{RVM} 
		& 0.070 & 47.377 & 0.091 & 0.560 & 0.001 & {0.976} \\
		& \textcolor{red}{MCMC-CSS} 
		& {0.028} & {4.101} & 0.080 & {0.051} & {0.778} & 0.972 \\
		& \textcolor{violet}{MCMC-DSS-i} 
		& {0.030} & {5.130} & {0.079} & {0.040} & {0.842} & 0.977 \\ 
		& \textcolor{blue}{MCMC-DSS-g} 
		& \textbf{0.023} & \textbf{3.761} & \textbf{0.077} & \textbf{0.026} & \textbf{0.898} & \textbf{0.977} \\ \midrule            
		%
		\multirow{4}{*}{\begin{tabular}[c]{@{}c@{}}Quadratic\\ damping \end{tabular}} 
		%
		& \textcolor{green!50!black!90!}{RVM} 
		& 0.017 & 1546.542 & 0.073 & 0.497 & 0.003 & \textbf{0.931} \\
		& \textcolor{red}{MCMC-CSS} 
		& \textbf{0.017} & {0.005} & 0.072 & \textbf{0.006} & \textbf{0.876} & 0.886 \\
		& \textcolor{violet}{MCMC-DSS-i} 
		& {0.018} & \textbf{0.004} & \textbf{0.072} & 0.031 & 0.850 & 0.855 \\    
		& \textcolor{blue}{MCMC-DSS-g} 
		& 0.020 & {0.004} & 0.072 & 0.031 & 0.845 & 0.847 \\     
		%
		\midrule
		\multirow{4}{*}{\begin{tabular}[c]{@{}c@{}}Coulomb\\ damping \end{tabular}} 
		%
		& \textcolor{green!50!black!90!}{RVM} 
		& 0.013 & 1034.780 & 0.092 & 0.496 & 0.003 & \textbf{0.993} \\
		& \textcolor{red}{MCMC-CSS} 
		& 0.010 & 1.476 & 0.071 & \textbf{0.002} & 0.769 & 0.775 \\
		& \textcolor{violet}{MCMC-DSS-i} 
		& 0.011 & 0.004 & 0.071 & 0.018 & \textbf{0.835} & 0.838 \\
		& \textcolor{blue}{MCMC-DSS-g} 
		& \textbf{0.009} & \textbf{0.004} & \textbf{0.070} & {0.010} & 0.838 & 0.840 \\  
		\bottomrule[1.5pt]
	\end{tabular}
	\caption{Comparison of results from RVM, MCMC-CSS, MCMC-DSS-i and MCMC-DSS-g, averaged over 1000 realisations.
		Small values of $e_{\theta}^s$, $e_{\theta}$, $e_{p}$, FDR are better, whereas average values of $\hat{\mathcal{M}}=\mathcal{M}$ and $\hat{\mathcal{M}} \supset \mathcal{M}$ closer to one are better; bold numbers highlight the best performing metric.}
	\label{table:compareMCS_A}	
\end{table}

Table \ref{table:compareMCS_A} shows that all the proposed MCMC-SS algorithms outperform the RVM in all metrics of performance. The MCMC-SS algorithms yield strikingly low levels of parameter estimation errors and false discoveries compared to the RVM. The RVM includes a lot of false discoveries, which is a major deterrent in equation discovery, as selecting the correct set of basis variables is crucial for drawing scientific conclusions based on the estimated model. Moreover, the models estimated by MCMC-SS surpass those by RVM in terms of predictive accuracy as well. Among the three variants of MCMC-SS algorithm, the DSS priors yield quite similar results and often tend to perform slightly better than the CSS prior. The RVM, however, is found to do better in superset model selection rate for the cases of quadratic viscous damping and Coulomb friction damping. In those cases, the RVM is able to include all the relevant variables in the estimated model more often than the MCMC-SS algorithms. The weaker sparsity-promoting property of the RVM allows it more often include all the relevant variables but with many other false discoveries.  


Overall, the MCMC-SS algorithms show very strong model selection consistency; they are able to select the true models more often and show extremely low rates of false discoveries -- an important requirement for intepretability of discovered equations. It can be inferred that the RVM (using the Student's-$t$ prior) very rarely finds the exact true model and will likely include many false discoveries. That being said, the RVM is remarkably fast compared to the MCMC-SS algorithms. A comparison of the average runtimes of the four sparse Bayesian learning algorithms are provided in Table \ref{table:avgRuntime}; the algorithms are run on a 64-bit Windows 10 PC with Intel Xeon E5-2698v4 CPU @ 2.20GHz. The RVM is undoubtedly the cheapest in terms of computational time, while all the MCMC-SS algorithms are orders of magnitude more expensive than the RVM. Note that the MCMC algorithms could be more time-consuming if the number of sampling iterations were increased. Between the three SS prior variants, the MCMC algorithm implemented with CSS priors is much cheaper than the DSS priors; the increased computational time for the DSS priors is due to the calculation of the marginal likelihood in Eq.~\eqref{eq:add_step_DSS}, needed for integrating out the Dirac-delta function.
\begin{table}[H]
	\centering
	\ra{1.1}
	\begin{tabular}{ccccccc} \toprule[1.5pt]
		\textcolor{green!50!black!90!}{RVM}   && \textcolor{red}{MCMC-CSS} && \textcolor{violet}{MCMC-DSS-i}  && \textcolor{blue}{MCMC-DSS-g} \\ \midrule[1pt]
		0.03s && 3.89s && 36.22s && 34.59s\\ 
		\bottomrule[1.5pt]
	\end{tabular}
	\caption{Average computational runtimes of RVM and MCMC-SS (run with single chain for 5000 sampling iterations).}
	\label{table:avgRuntime}
\end{table}  


\section{An experimental application on Silverbox benchmark} \label{sec:exp_Silverbox}
This section presents an application of SS priors for Bayesian equation discovery of the Silverbox benchmark \cite{wigren2013three,wigren2013data}. The Silverbox is an electrical circuit resembling a Duffing oscillator with a moving mass $m$, a viscous damping $c$ and a nonlinear spring $k(q)$. The circuit is designed to relate the displacement $q(t)$ (the output) to the force $u(t)$ (the input) by the following differential equation,
\begin{equation}
	m \ddot{q}(t) + c\dot{q} (t) + \underbrace{\brc{a + b q^2(t)}}_{k(q(t))} q(t) = u(t)
\end{equation}
The input-output data for the Silverbox benchmark consist of force and displacement measurements. The data here were supplied as part of the \textit{Nonlinear System Identification Benchmarks} workshop held at VUB Brussels and Eindhoven University over the last few years. More details on the experiment and benchmark data can be found in \cite{wigren2013three,wigren2013data}. 

\begin{figure}[h!]
	\begin{center}
		\includegraphics[scale=0.75]{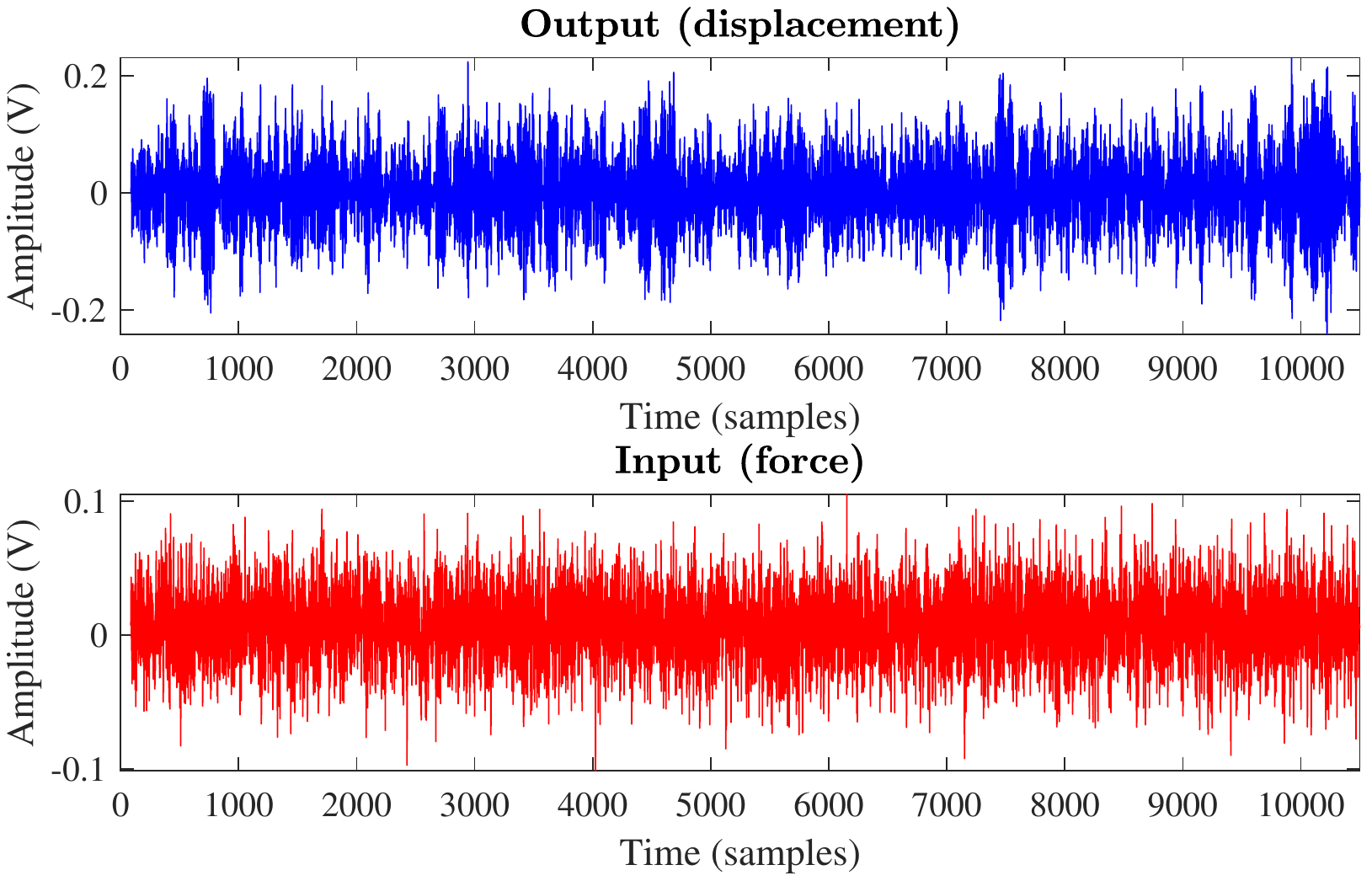}
		\caption{The portion of input-output measurement data from Silverbox that was used in training the sparse Bayesian algorithms; the input consists of random-phase multi-sine excitation.}
		\label{f:trainIO_data_silverbox}
	\end{center}
\end{figure}

The \texttt{Schroeder80mV.mat} dataset of the Silverbox benchmark has been used here for the application. A section of the input-output data consisting of 10400 samples (between 0.14s and 17.2s), as shown in Figure~\ref{f:trainIO_data_silverbox}, was used for training the sparse Bayesian algorithms; the input data comprised a random-phase multi-sine excitation containing 1342 odd harmonics of a base frequency ${8192}/{f_{\text{samp}}} \si{Hz}$ and the sampling rate $f_{\text{samp}}$ in the experiments was $610.35\si{Hz}$. Since the displacement and force signals were the only data measured, numerical differentiation was employed to obtain the velocity and acceleration signals from the displacement data. This case, therefore, serves as a measurement scenario where only one output response is observed. 

The results of parameter estimation from MCMC-SS and RVM algorithms are presented in Table~\ref{table:Silverbox_PE_results}. The MCMC-SS algorithms selected five basis variables which include the correct linear stiffness term $x_1$, the viscous damping $x_2$, the nonlinear cubic stiffness  $x_1^3$, the input force $u$, and a spurious term $3x_1 x_2^2$. A comparison of the posterior means ($\hat{\mu}_{\vtheta}$) of the weights from the MCMC-SS algorithms reveals that the cubic term $x_1^3$ is the dominant term followed by $x_1$, $u$ and $x_2$, while the weight associated with the spurious term $3x_1 x_2^2$ is quite small. 
As observed previously, the RVM selected many more basis variables apart from the set of basis variables deemed relevant by the MCMC-SS algorithms. However, most of the spurious terms are associated with large standard deviations compared to their mean values, and may be discarded based on the degree of uncertainty. For example, the spurious term $x_1^2$ selected by RVM has a posterior standard deviation comparable to its posterior mean, and can be ignored.

\begin{table}[h!] 
	\centering
	\ra{1.2}
	\begin{tabular}{c c c c  c}
		\toprule[1.5pt]
		\multirow{2}{*}{\begin{tabular}[c]{@{}l@{}}Relevant\\ variables\end{tabular}} & \multicolumn{4}{c}{Estimated mean and standard derivations of unscaled weights ($\hat{\mu}_{\vtheta} \pm \hat{\sigma}_{\vtheta}$)}  \\ \cmidrule{2-5}
		& \textcolor{green!50!black!90!}{RVM} & \textcolor{red}{MCMC-CSS}  & \textcolor{violet}{MCMC-DSS-i}  & \textcolor{blue}{MCMC-DSS-g}  \\ \midrule[1pt]
		{$-x_1$} & $(14.99 \pm 0.09) \times 10^4$ & $(15.03 \pm 0.11) \times 10^4$ & $(15.10 \pm 0.04)\times 10^4$ &  $(15.10 \pm 0.03)\times 10^4$\\
		{$-x_2$} & $26.27 \pm 0.77$ & $25.68 \pm 1.24$ & $25.32 \pm 0.46$ & $25.31 \pm 0.45$ \\
		$-2 x_1 x_2$ & $2.31 \pm 3.86$ & 0 & 0 & 0 \\
		$-x_1^2$ & $-3185.04 \pm 1833.56$ & 0 & 0 & 0 \\
		$-3 x_1 x_2^2$ & $0.94 \pm 0.17$ & $0.72 \pm 0.23$ & $0.69 \pm 0.11$ & $0.69 \pm 0.11$ \\ 
		$-3 x_1^2 x_2$ & $-33.04 \pm 24.56$ & 0 & 0 & 0 \\ 
		{$-x_1^3$} & $(35.38 \pm 7.33) \times 10^4$ & $(37.42 \pm 10.55) \times 10^4$ & $(40.50 \pm 2.47) \times 10^4$ & $(40.58 \pm 2.03)\times 10^4$ \\ 
		$-30x_1 x_2^2$ & $-2.57 \pm 2.27$ & 0 & 0 & 0 \\ 
		$-6x_1^5x_2$ & $-2696.45 \pm 3008.18$ & 0 & 0 & 0 \\ 
		$-\text{sgn} (x_2)$ & $-17.71 \pm 17.76$ & 0 & 0 & 0 \\ 
		$-x_1\abs{x_1}$ & $(15.76 \pm 16.56) \times 10^4$ & 0 & 0 & 0 \\ 
		{$u$} & $(10.17 \pm 0.04) \times 10^4$ & $(10.17 \pm 0.04)\times 10^4$ & $(10.17 \pm 0.04)\times 10^4$ & $(10.17 \pm 0.04)\times 10^4$ \\ 
		\bottomrule[1.5pt]
	\end{tabular}
	\caption{Variable selection and parameter estimation results for the Silverbox nonlinear benchmark using RVM and MCMC-SS algorithms; leftmost column enumerates the set of basis variables deemed relevant by at least one of the four sparse Bayesian algorithms, and the rest of columns show the posterior means and standard deviations of the estimated unscaled weights.}
	\label{table:Silverbox_PE_results}	
\end{table}

\begin{figure}[h!]
	\begin{center}
		\includegraphics[scale=0.80]{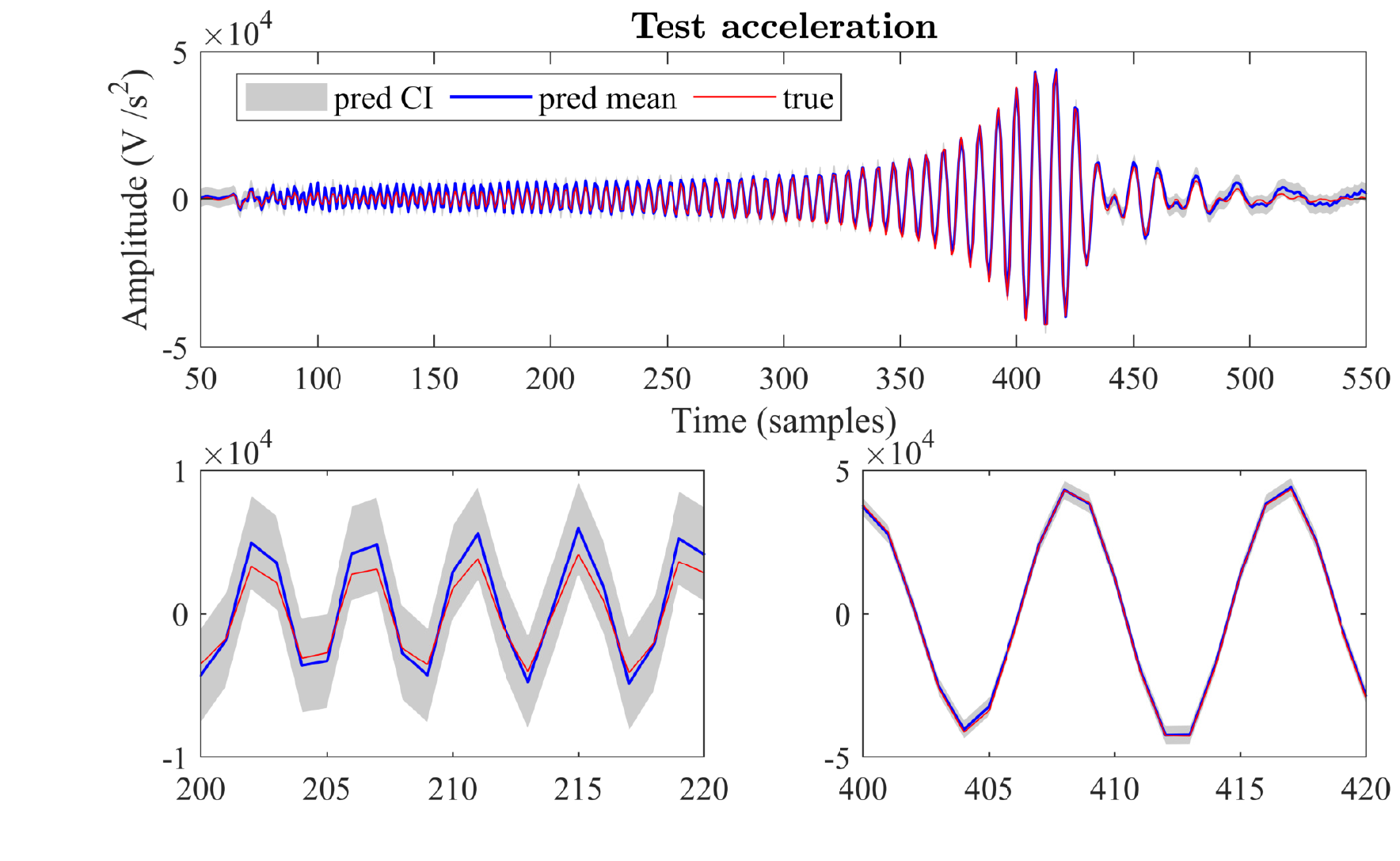}
		\caption{Test set prediction results on the Silverbox nonlinear benchmark from MCMC-DSS-i algorithm; testing with down-chirp input excitation. The top figure along with the ``zoomed-in'' bottom figures show the predicted mean and $3\sigma$ confidence interval (CI) of the test acceleration plotted alongside with `true' acceleration. Predictions are better at lower input frequencies (bottom-right figure) than at higher input frequencies (bottom-left figure). }
		\label{f:testset_accpred_silverbox}
	\end{center}
\end{figure}

To assess the predictive power of the discovered models, an independent input-output dataset -- where the input excitation is a chirp signal going from high to low frequencies -- was used for testing. The test input-output signals correspond to a set of force-displacement samples from the \texttt{Schroeder80mV.mat} dataset, and the `true' test acceleration was obtained by numerical differentiation of the measured displacement data. It was found that the test set prediction results were very similar for all the four sparse Bayesian algorithms, with the prediction errors being $0.132$ for RVM and $0.131$ for the three MCMC-SS algorithms. To illustrate the prediction performance, the result from the MCMC-DSS-i algorithm is used as a representative, and the predicted mean and confidence intervals (CIs) of the test acceleration signal from the algorithm are plotted alongside the `true' test acceleration signal in Figure~\ref{f:testset_accpred_silverbox}. The prediction results show an excellent match around resonance (bottom-right subplot of Figure~\ref{f:testset_accpred_silverbox}), which occurs at lower frequencies of the down-chirp input. However, some discrepancies are seen at higher input frequencies (bottom-left subplot); nonetheless, the `true' values are always captured by the predicted CIs.


\section{Discussion} \label{sec:discussion}
From the standpoint of equation discovery, the results of Bayesian variable selection and parameter estimation using the MCMC-SS algorithms are quite encouraging. The estimated models (or governing equations) are not only more interpretable but also superior in prediction, compared to the RVM. However, unlike RVM, the MCMC-SS algorithms are based on random sampling and are orders of magnitude more expensive than the RVM. For faster implementation of Bayesian inference with SS priors, one may consider employing alternative methods such as expectation maximisation \cite{rockova2014emvs}, variational Bayes \cite{titsias2011spike,carbonetto2012scalable,ormerod2017variational}, expectation propagation \cite{hernandez2015expectation}, etc. 

The dictionary of candidate basis variables plays a significant role in practical implementation of equation discovery algorithms. The success in discovering correct equations greatly hinges on whether or not the true basis variables are included in the dictionary. Absence of the true variables in the dictionary will lead to discoveries of terms that are strongly correlated with the true variables. For example, if $\sin(x)$ is a true variable that is not included in the dictionary, one will end up selecting correlated variables such as $x, x^3, x^5, \ldots$ from a polynomial-based dictionary. Ideally, the dictionary should consist of as many diverse bases as possible, not just polynomial bases. In practice, however, putting more basis variables in the dictionary increases the correlation between basis variables and causes ill-conditioning of the dictionary matrix. The DSS-g prior was introduced to account for the correlation structure of the bases in the dictionary; its use was found to result in reduced uncertainty in the parameter estimates, however, the mean values of the parameter estimates was not found to be significantly different from that of the DSS-i prior. It was also found that a severely ill-conditioned dictionary can cause poor mixing (or even local entrapment) of the Markov chains and can induce incorrect selection of variables from a set of correlated basis variables.  
Parallel-tempering algorithms \cite{earl2005parallel} can aid in improving the mixing of chains; however, they would significantly increase the computational burden. A reasonable approach would be to assess the dictionary for strongly-correlated basis variables prior to Bayesian inference, and if possible, eliminate some of them after careful deliberation. For example, variables $x$ and $\sin(x)$ are highly correlated for small values of $x$, and one may choose to exclude $\sin(x)$ from the dictionary to prevent ill-conditioning, as has been done here. In the experience of the authors, the RVM is more robust in handling ill-conditioned dictionaries than MCMC-SS algorithms. 


It should also be mentioned that the accuracy of the equation discovery approach can be greatly affected by errors in the state variables $\x$, present either in the form of measurement noise or state estimation errors. Since the basis variables are dependent on the states $\x$, even moderate amounts of errors in $\x$ can nonlinearly corrupt the constructed bases in the dictionary, and will eventually result in discovering incorrect equations. In the numerical study in Section \ref{sec:numerical}, measurements of all three response variables, i.e., displacement, velocity and acceleration were assumed, and a relatively small noise was used to corrupt the measurements. While all three variables can be measured for a small system in a laboratory setup, it would be prohibitive to do so in practice, especially for large structural systems. 
A truly pragmatic approach would be to measure acceleration -- the most commonly measured quantity in structural testing -- and estimate the displacement and velocity from them. However, a naive numerical integration of the acceleration to obtain displacement or velocity may not work well, as even small amounts of noise in the displacement or velocity could deteriorate the equation discovery results. 
Recently, a promising optimisation framework was proposed in \cite{kaheman2020automatic}, that leverages automatic differentiation and sparse regression to simultaneously separate the noise signal from the measured data as well as recover the governing differential equations via numerical time-stepping constraints. This approach could be combined with a sparse Bayesian learning framework to make the equation discovery procedure more robust. Future efforts will look at developing robust approaches of equation discovery using acceleration measurements.

\section{Conclusions} \label{sec:conclusions}
This paper presents a novel application of SS priors in Bayesian equation discovery of structural dynamic systems, which aims at discovering the governing ordinary differential equations of motion of a structural system from measured input-output data. The equation discovery procedure is tantamount to a simultaneous model selection and parameter estimation problem in system identification. Using a dictionary of nonlinear bases variables composed using the measured data, the problem of Bayesian model selection is turned into a Bayesian variable selection problem and solved via sparse linear regression, thus bypassing a combinatorially large search through all possible candidate models. The SS priors are well-known to possess superior sparsity-enforcing properties compared to the Laplace or Student's-$t$ priors, owing to their two-component mixture distributions of a narrow spike and a comparably flat slab. As such, their use in variable selection has the potential to derive more parsimonious and interpretable equations of motion. 
In this paper, three different variants of SS priors -- namely the CSS, DSS-i and DSS-g -- are employed as prior distributions, and MCMC-based Gibbs sampling algorithms are derived to select the relevant variables and estimate associated parameters. 


Using a series of numerical simulations, it has been demonstrated that the proposed MCMC-SS algorithms correctly identify the presence and type of various nonlinearities such as a cubic stiffness, a quadratic viscous damping, and a Coulomb friction damping. Furthermore, using Monte Carlo simulations, the performance of MCMC-SS has been compared to RVM, which uses the Student's-$t$ prior. It is found that MCMC-SS algorithms display stronger model selection consistency than the RVM. Additionally, the predictive accuracy of the models selected by MCMC-SS is found to be highly competitive to that of the models estimated by RVM.

\section{Acknowledgements}
This work has been funded by the UK Engineering and Physical Sciences Research Council (EPSRC), via the Autonomous Inspection in Manufacturing and Re-manufacturing (AIMaReM) grant EP/N018427/1. Support for K.\ Worden from the EPSRC via grant reference number EP/J016942/1 and for E.J.\ Cross through grant number EP/S001565/1 is also gratefully acknowledged. 

\appendix
\section{Gibbs sampling scheme for CSS prior} \label{sec:MCMCsteps_CSS}
The Gibbs sampling steps for the CSS prior are adopted from the BASAD algorithm \cite{narisetty2014bayesian}. Note that, as mentioned in Section \ref{sec:SSmodel}, the CSS priors involve a vector of slab variances, that is, a slab variance $v_{s_i}$ is associated with each weight $\theta_i$. The steps for the parameters $\vtheta$, $\z$, $\pz$, $v_{s_i}$ and $\sigma^2$ are as follows:
\begin{enumerate} [label=(\alph*)]
	\item $\vtheta$ is sampled from a Gaussian distribution as follows,
	\begin{align} \label{eq:sampleW_CSS}
		\vtheta \mid \y, \z, \vs, \sigma^2 \sim \ND{\muz^{}, \sigma^2 \Sigmaz^{}}
	\end{align}
	where $\Sigmaz^{} = \brc{\mD^{T} \mD + \matr{V}^{-1}}^{-1}$, $\muz^{} = \Sigmaz^{} \mD^{T} \y$, and $\matr{V}$ is a diagonal matrix with  elements $\matr{V}_{i,i} = v_{s_i} \brc{v_0(1-z_i) + v_1 z_i}$, $i=1,\ldots,P$.
	\item $\sigma^2$ is sampled from an inverse Gamma distribution as follows,
	\begin{align} \label{eq:sampleSig_CSS}
		\sigma^2 \mid \y, \vtheta \sim \IG{a_{\sigma} + \frac{N}{2} + \frac{P}{2}, \; b_{\sigma} + \frac{1}{2} \brc{\brc{\y-\mD \vtheta}^T \brc{\y-\mD \vtheta} + \vtheta^T \matr{V}^{-1} \vtheta} }
	\end{align}
	\item The vector of weight-specific slab variances is sampled componentwise. The $i^{\text{th}}$ component, $v_{s_i}$, is sampled from an inverse Gamma distribution as follows,
	\begin{align} \label{eq:sampleVs_CSS}
		v_{s_i} \mid \vtheta, \z \sim \IG{a_v + \frac{1}{2}, \; b_v +  \frac{0.5\theta_i^2}{2 \sigma^2 \brc{v_0(1-z_i) + v_1 z_i}} }
	\end{align}
	\item $\pz$ is sampled from a Beta distribution as follows,
	\begin{align} \label{eq:sampleP_CSS}
		\pz \mid \z \sim \text{Beta} \brc{a_p + \sum_{i=1}^P z_i, \; b_p + P - \sum_{i=1}^P z_i}
	\end{align}
	\item The conditional distribution of $\z$ is expressed componentwise. The odds of $z_i= 1$ to $z_i = 0$ are computed. The components of $\z$ are sampled as follows, 
	\begin{align} \label{eq:sampleZ_CSS}
		z_i \mid \theta_i, v_{s_i}, \sigma^2 &\sim \text{Bern}(\xi_i),  \text{   with } \xi_i = \frac{\pz}{\pz + \frac{\pr{\theta_i \mid z_i = 0, v_{s_i}, \sigma^2} }{\pr{\theta_i \mid z_i = 1, v_{s_i}, \sigma^2} }  (1-\pz)}
	\end{align}
	In the above sampling step, the probabilities $\pr{\theta_i \mid z_i=0, v_{s_i}, \sigma^2}$ and $\pr{\theta_i \mid z_i=1, v_{s_i}, \sigma^2}$ can be computed by evaluating the Gaussian densities over $\theta_i$, as follows:
	\begin{align*}
		\pr{\theta_i \mid z_i=0, v_{s_i}, \sigma^2} &= \frac{1}{\sqrt{2 \pi \sigma^2 v_0 v_{s_i}}} \exp \brc{-\frac{\theta_i^2}{2 \sigma^2 v_0 v_{s_i}}}\\
		\pr{\theta_i \mid z_i=1, v_{s_i}, \sigma^2} &= \frac{1}{\sqrt{2 \pi \sigma^2 v_1 v_{s_i}}} \exp \brc{-\frac{\theta_i^2}{2 \sigma^2 v_1 v_{s_i}}}
	\end{align*}
\end{enumerate}

\bibliographystyle{unsrt}  
\bibliography{References}

\begin{thebibliography}{10}

\bibitem{bongard2007automated}
Josh Bongard and Hod Lipson.
\newblock Automated reverse engineering of nonlinear dynamical systems.
\newblock {\em Proceedings of the National Academy of Sciences},
  104(24):9943--9948, 2007.

\bibitem{schmidt2009distilling}
Michael Schmidt and Hod Lipson.
\newblock Distilling free-form natural laws from experimental data.
\newblock {\em Science}, 324(5923):81--85, 2009.

\bibitem{brunton2016discovering}
Steven~L Brunton, Joshua~L Proctor, and J~Nathan Kutz.
\newblock Discovering governing equations from data by sparse identification of
  nonlinear dynamical systems.
\newblock {\em Proceedings of the National Academy of Sciences},
  113(15):3932--3937, 2016.

\bibitem{mangan2016inferring}
Niall~M Mangan, Steven~L Brunton, Joshua~L Proctor, and J~Nathan Kutz.
\newblock Inferring biological networks by sparse identification of nonlinear
  dynamics.
\newblock {\em IEEE Transactions on Molecular, Biological and Multi-Scale
  Communications}, 2(1):52--63, 2016.

\bibitem{schaeffer2017sparse}
Hayden Schaeffer and Scott~G McCalla.
\newblock Sparse model selection via integral terms.
\newblock {\em Physical Review E}, 96(2):023302, 2017.

\bibitem{mangan2017model}
Niall~M Mangan, J~Nathan Kutz, Steven~L Brunton, and Joshua~L Proctor.
\newblock Model selection for dynamical systems via sparse regression and
  information criteria.
\newblock {\em Proceedings of the Royal Society A: Mathematical, Physical and
  Engineering Sciences}, 473(2204):20170009, 2017.

\bibitem{kaiser2018sparse}
Eurika Kaiser, J~Nathan Kutz, and Steven~L Brunton.
\newblock Sparse identification of nonlinear dynamics for model predictive
  control in the low-data limit.
\newblock {\em Proceedings of the Royal Society A}, 474(2219):20180335, 2018.

\bibitem{champion2019data}
Kathleen Champion, Bethany Lusch, J~Nathan Kutz, and Steven~L Brunton.
\newblock Data-driven discovery of coordinates and governing equations.
\newblock {\em Proceedings of the National Academy of Sciences},
  116(45):22445--22451, 2019.

\bibitem{schaeffer2020extracting}
Hayden Schaeffer, Giang Tran, Rachel Ward, and Linan Zhang.
\newblock Extracting structured dynamical systems using sparse optimization
  with very few samples.
\newblock {\em Multiscale Modeling \& Simulation}, 18(4):1435--1461, 2020.

\bibitem{boninsegna2018sparse}
Lorenzo Boninsegna, Feliks N{\"u}ske, and Cecilia Clementi.
\newblock Sparse learning of stochastic dynamical equations.
\newblock {\em The Journal of Chemical Physics}, 148(24):241723, 2018.

\bibitem{mangan2019model}
Niall~M Mangan, Travis Askham, Steven~L Brunton, J~Nathan Kutz, and Joshua~L
  Proctor.
\newblock Model selection for hybrid dynamical systems via sparse regression.
\newblock {\em Proceedings of the Royal Society A}, 475(2223):20180534, 2019.

\bibitem{stender2019recovery}
Merten Stender, Sebastian Oberst, and Norbert Hoffmann.
\newblock Recovery of differential equations from impulse response time series
  data for model identification and feature extraction.
\newblock {\em Vibration}, 2(1):25--46, 2019.

\bibitem{rudy2017data}
Samuel~H Rudy, Steven~L Brunton, Joshua~L Proctor, and J~Nathan Kutz.
\newblock Data-driven discovery of partial differential equations.
\newblock {\em Science Advances}, 3(4):e1602614, 2017.

\bibitem{zhang2018robust}
Sheng Zhang and Guang Lin.
\newblock Robust data-driven discovery of governing physical laws with error
  bars.
\newblock {\em Proceedings of the Royal Society A: Mathematical, Physical and
  Engineering Sciences}, 474(2217):20180305, 2018.

\bibitem{chen2020deep}
Zhao Chen, Yang Liu, and Hao Sun.
\newblock Deep learning of physical laws from scarce data.
\newblock {\em arXiv preprint arXiv:2005.03448}, 2020.

\bibitem{rudy2019deep}
Samuel~H Rudy, J~Nathan Kutz, and Steven~L Brunton.
\newblock Deep learning of dynamics and signal-noise decomposition with
  time-stepping constraints.
\newblock {\em Journal of Computational Physics}, 396:483--506, 2019.

\bibitem{raissi2018multistep}
Maziar Raissi, Paris Perdikaris, and George~Em Karniadakis.
\newblock Multistep neural networks for data-driven discovery of nonlinear
  dynamical systems.
\newblock {\em arXiv preprint arXiv:1801.01236}, 2018.

\bibitem{raissi2018deep}
Maziar Raissi.
\newblock Deep hidden physics models: {D}eep learning of nonlinear partial
  differential equations.
\newblock {\em The Journal of Machine Learning Research}, 19(1):932--955, 2018.

\bibitem{kerschen2006past}
Gaetan Kerschen, Keith Worden, Alexander~F Vakakis, and Jean-Claude Golinval.
\newblock Past, present and future of nonlinear system identification in
  structural dynamics.
\newblock {\em Mechanical Systems and Signal Processing}, 20(3):505--592, 2006.

\bibitem{noel2017nonlinear}
Jean-Philippe No{\"e}l and Ga{\"e}tan Kerschen.
\newblock Nonlinear system identification in structural dynamics: 10 more years
  of progress.
\newblock {\em Mechanical Systems and Signal Processing}, 83:2--35, 2017.

\bibitem{akaike1974new}
Hirotugu Akaike.
\newblock A new look at the statistical model identification.
\newblock {\em IEEE Transactions on Automatic Control}, 19(6):716--723, 1974.

\bibitem{schwarz1978estimating}
Gideon~E. Schwarz.
\newblock Estimating the dimension of a model.
\newblock {\em The Annals of Statistics}, 6(2):461--464, 1978.

\bibitem{hastie2015statistical}
Trevor Hastie, Robert Tibshirani, and Martin Wainwright.
\newblock {\em Statistical {L}earning with {S}parsity: the {L}asso and
  {G}eneralizations}.
\newblock CRC press, 2015.

\bibitem{tipping2001sparse}
Michael~E Tipping.
\newblock Sparse {B}ayesian learning and the relevance vector machine.
\newblock {\em Journal of Machine Learning Research}, 1:211--244, 2001.

\bibitem{wipf2004sparse}
David~P Wipf and Bhaskar~D Rao.
\newblock Sparse {B}ayesian learning for basis selection.
\newblock {\em IEEE Transactions on Signal processing}, 52(8):2153--2164, 2004.

\bibitem{seeger2010optimization}
Matthias~W Seeger, Hannes Nickisch, Rolf Pohmann, and Bernhard Sch{\"o}lkopf.
\newblock Optimization of k-space trajectories for compressed sensing by
  {B}ayesian experimental design.
\newblock {\em Magnetic Resonance in Medicine: An Official Journal of the
  International Society for Magnetic Resonance in Medicine}, 63(1):116--126,
  2010.

\bibitem{seeger2008bayesian}
Matthias~W Seeger.
\newblock Bayesian inference and optimal design for the sparse linear model.
\newblock {\em Journal of Machine Learning Research}, 9:759--813, 2008.

\bibitem{carvalho2009handling}
Carlos~M Carvalho, Nicholas~G Polson, and James~G Scott.
\newblock Handling sparsity via the horseshoe.
\newblock In {\em Artificial Intelligence and Statistics}, pages 73--80, 2009.

\bibitem{mitchell1988bayesian}
Toby~J Mitchell and John~J Beauchamp.
\newblock Bayesian variable selection in linear regression.
\newblock {\em Journal of the American Statistical Association},
  83(404):1023--1032, 1988.

\bibitem{george1993variable}
Edward~I George and Robert~E McCulloch.
\newblock Variable selection via {G}ibbs sampling.
\newblock {\em Journal of the American Statistical Association},
  88(423):881--889, 1993.

\bibitem{geweke1996variable}
John Geweke.
\newblock Variable selection and model comparison in regression.
\newblock {\em In Bayesian Statistics 5: Proceedings of the Fifth Valencia
  International Meeting}, pages 609--620, 1996.

\bibitem{george1997approaches}
Edward~I George and Robert~E McCulloch.
\newblock Approaches for {B}ayesian variable selection.
\newblock {\em Statistica Sinica}, pages 339--373, 1997.

\bibitem{ishwaran2005spike}
Hemant Ishwaran and J~Sunil Rao.
\newblock Spike and slab variable selection: {F}requentist and {B}ayesian
  strategies.
\newblock {\em The Annals of Statistics}, 33(2):730--773, 2005.

\bibitem{polson2010shrink}
Nicholas~G Polson and James~G Scott.
\newblock Shrink globally, act locally: {S}parse {B}ayesian regularization and
  prediction.
\newblock {\em Bayesian statistics}, 9(501-538):105, 2010.

\bibitem{van2019shrinkage}
Sara van Erp, Daniel~L Oberski, and Joris Mulder.
\newblock Shrinkage priors for {B}ayesian penalized regression.
\newblock {\em Journal of Mathematical Psychology}, 89:31--50, 2019.

\bibitem{fuentes2019efficient}
R~Fuentes, N~Dervilis, K~Worden, and Elizabeth~J Cross.
\newblock Efficient parameter identification and model selection in nonlinear
  dynamical systems via sparse {B}ayesian learning.
\newblock In {\em Journal of Physics: Conference Series}, volume 1264, page
  012050. IOP Publishing, 2019.

\bibitem{zhang2019robust}
Sheng Zhang and Guang Lin.
\newblock Robust data-driven discovery of governing physical laws using a new
  subsampling-based sparse {B}ayesian method to tackle four challenges (large
  noise, outliers, data integration, and extrapolation).
\newblock {\em arXiv preprint arXiv:1907.07788}, 2019.

\bibitem{nayek2020spike}
Rajdip Nayek, Keith Worden, Elizabeth~J Cross, and Ramon Fuentes.
\newblock A sparse {B}ayesian approach to model structure selection and
  parameter estimation of dynamical systems using spike-and-slab priors.
\newblock In {\em Proceedings of the 8th International Conference on Noise and
  Vibration Engineering (ISMA2020)}, Sep 2020.

\bibitem{ohara2009review}
Robert~B O'Hara and Mikko~J Sillanp{\"a}{\"a}.
\newblock A review of {B}ayesian variable selection methods: what, how and
  which.
\newblock {\em Bayesian analysis}, 4(1):85--117, 2009.

\bibitem{narisetty2014bayesian}
Naveen~Naidu Narisetty and Xuming He.
\newblock Bayesian variable selection with shrinking and diffusing priors.
\newblock {\em The Annals of Statistics}, 42(2):789--817, 2014.

\bibitem{liang2008mixtures}
Feng Liang, Rui Paulo, German Molina, Merlise~A Clyde, and Jim~O Berger.
\newblock Mixtures of g-priors for {B}ayesian variable selection.
\newblock {\em Journal of the American Statistical Association},
  103(481):410--423, 2008.

\bibitem{andrews1974scale}
David~F Andrews and Colin~L Mallows.
\newblock Scale mixtures of normal distributions.
\newblock {\em Journal of the Royal Statistical Society: Series B
  (Methodological)}, 36(1):99--102, 1974.

\bibitem{malsiner2011comparing}
Gertraud Malsiner-Walli and Helga Wagner.
\newblock Comparing spike and slab priors for {B}ayesian variable selection.
\newblock {\em Austrian Journal of Statistics}, 40(4):241--264, 2011.

\bibitem{casella1992explaining}
George Casella and Edward~I George.
\newblock Explaining the {G}ibbs sampler.
\newblock {\em The American Statistician}, 46(3):167--174, 1992.

\bibitem{barbieri2004optimal}
Maria~Maddalena Barbieri and James~O Berger.
\newblock Optimal predictive model selection.
\newblock {\em The Annals of Statistics}, 32(3):870--897, 2004.

\bibitem{worden2019nonlinearity}
Keith Worden and G.~R. Tomlinson.
\newblock {\em Nonlinearity in {S}tructural {D}ynamics: {D}etection,
  {I}dentification and {M}odelling}.
\newblock CRC Press, 2019.

\bibitem{dahl1976solid}
Philip~R Dahl.
\newblock Solid friction damping of mechanical vibrations.
\newblock {\em AIAA Journal}, 14(12):1675--1682, 1976.

\bibitem{brooks1998general}
Stephen~P Brooks and Andrew Gelman.
\newblock General methods for monitoring convergence of iterative simulations.
\newblock {\em Journal of computational and graphical statistics},
  7(4):434--455, 1998.

\bibitem{tipping2003fast}
Michael~E Tipping and Anita~C Faul.
\newblock Fast marginal likelihood maximisation for sparse {B}ayesian models.
\newblock In {\em Proceedings of the Ninth AISTATS Conference}, pages 1--13,
  2003.

\bibitem{RVMsoftware}
{SparseBayes Software} v2.0.
\newblock \url{http://www.miketipping.com/downloads.htm}.
\newblock {L}ast Accessed: 2020-11-10.

\bibitem{wigren2013three}
Torbj{\"o}rn Wigren and Johan Schoukens.
\newblock Three free data sets for development and benchmarking in nonlinear
  system identification.
\newblock In {\em 2013 European control conference (ECC)}, pages 2933--2938.
  IEEE, 2013.

\bibitem{wigren2013data}
Torbj{\"o}rn Wigren and Johan Schoukens.
\newblock Data for benchmarking in nonlinear system identification.
\newblock {\em Technical Reports from the department of Information
  Technology}, 6:2013--006, 2013.

\bibitem{rockova2014emvs}
Veronika Ro{\v{c}}kov{\'a} and Edward~I George.
\newblock {EMVS}: The {EM} approach to {B}ayesian variable selection.
\newblock {\em Journal of the American Statistical Association},
  109(506):828--846, 2014.

\bibitem{titsias2011spike}
Michalis~K Titsias and Miguel L{\'a}zaro-Gredilla.
\newblock Spike and slab variational inference for multi-task and multiple
  kernel learning.
\newblock In {\em Advances in Neural Information Processing Systems}, pages
  2339--2347, 2011.

\bibitem{carbonetto2012scalable}
Peter Carbonetto and Matthew Stephens.
\newblock Scalable variational inference for {B}ayesian variable selection in
  regression, and its accuracy in genetic association studies.
\newblock {\em Bayesian Analysis}, 7(1):73--108, 2012.

\bibitem{ormerod2017variational}
John~T Ormerod, Chong You, and Samuel M{\"u}ller.
\newblock A variational {B}ayes approach to variable selection.
\newblock {\em Electronic Journal of Statistics}, 11(2):3549--3594, 2017.

\bibitem{hernandez2015expectation}
Jos{\'e}~Miguel Hern{\'a}ndez-Lobato, Daniel Hern{\'a}ndez-Lobato, and Alberto
  Su{\'a}rez.
\newblock Expectation propagation in linear regression models with
  spike-and-slab priors.
\newblock {\em Machine Learning}, 99(3):437--487, 2015.

\bibitem{earl2005parallel}
David~J Earl and Michael~W Deem.
\newblock Parallel tempering: {T}heory, applications, and new perspectives.
\newblock {\em Physical Chemistry Chemical Physics}, 7(23):3910--3916, 2005.

\bibitem{kaheman2020automatic}
Kadierdan Kaheman, Steven~L Brunton, and J~Nathan Kutz.
\newblock Automatic differentiation to simultaneously identify nonlinear
  dynamics and extract noise probability distributions from data.
\newblock {\em arXiv preprint arXiv:2009.08810}, 2020.

\end{thebibliography}


%
%
%
%

\end{document}